\documentclass[review, nonatbib]{elsarticle}

\makeatletter
\let\c@author\relax
\makeatother
\usepackage{hyperref}
\usepackage[nolist,nohyperlinks]{acronym}
\usepackage[style=numeric, defernumbers=true, backend=bibtex]{biblatex}
\graphicspath{{./images/}}
\usepackage{listings}
\usepackage{color}
\usepackage{tikz}
\usepackage{verbatim}
\usepackage{csvsimple,longtable,booktabs}
\usepackage[strict]{changepage}
\usepackage{caption}
\usepackage{float}
\usepackage{multirow}
\usepackage{multicol}
\usepackage{pbox}
\usepackage{enumitem}
\usepackage{selinput}	
\usepackage{todonotes}
\usepackage{hyperref}
\usepackage[titletoc]{appendix}
\usepackage{tabularx}
\usepackage{color, colortbl}
\definecolor{Gray}{gray}{0.9}
\let\textquotedbl="
\usepackage[autostyle=true]{csquotes}

\usepackage{setspace}
\usepackage{tabularx} 
\usepackage{rotating}
\usepackage{lscape}
\usepackage{longtable}
\usepackage{xltabular}
\usepackage{pifont}

\usepackage{placeins}

\usepackage{booktabs}

\usepackage{array}
\usepackage{longtable}

\hypersetup{breaklinks=true}

\journal{Journal of Systems and Software}

\begin{document}

\begin{acronym}
	\acro{AI} {Artificial Intelligence}
	\acro{AIOps}{Artificial Intelligence for IT Operations}
	\acro{API}{Application Programming Interface}
	\acro{AutoML}{Automated Machine Learning}
	\acro{CD}{Continuous Delivery}
	\acro{CD4ML}{Continuous Delivery for Machine Learning}
	\acro{CI}{Continuous Integration}
	\acro{CLEVER}{Cross Lipschitz Extreme Value for nEtwork Robustness}
	\acro{CPU}{Central Processing Unit}
	\acro{CRISP-DM}{CRoss-Industry Standard Process for Data Mining}
	\acro{CT}{Continuous Training}
	\acro{DAG}{Directed Acyclic Graph}
	\acro{DataOps}{Data and Operations}
	\acro{Dev}{Development}
	\acro{DevOps}{Development and Operations}
	\acro{DL}{Deep Learning}
	\acro{DSL}{Domain Specific Language}
	\acro{DVC}{Data Version Control}
	\acro{GPU}{Graphics Processing Unit}
	\acro{gRPC}{gRPC Remote Procedure Call}
	\acro{GUI}{Graphical User Interface}
	\acro{IoT}{Internet of Things}
    \acro{KPI}{Key Performance Indicator}
	\acro{LIME}{Local Interpretable Model Agnostic Explanations}
	\acro{ML}{Machine Learning}
	\acro{MLOps}{Machine Learning Operations}
	\acro{MLR}{Multivocal Literature Review}
	\acro{ModelOps}{Model and Operations}
	\acro{Ops}{Operations}
    \acro{PCA}{Principle Component Analysis}
	\acro{PDP}{Partial Dependency Plots}
	\acro{PFA}{Portable Format for Analytics}
	\acro{PMML}{Predictive Model Markup Language}
	\acro{QFD}{Quality Function Deployment}
	\acro{RAM}{Random Access Memory}
	\acro{RMSE}{Root Mean Square Error}
	\acro{ROC}{Receiver Operating Characteristic}
	\acro{SDK}{Software Developmen﻿t Kit}
	\acro{SEMMA}{Sample, Explore, Modify, Model and Assess}
	\acro{SHAP}{SHapley Additive exPlanations}
	\acro{SLR}{Systematic Literature Review}
    \acro{SMS}{Systematic Mapping Study}
	\acro{TDSP}{Team Data Science Process}
	\acro{TFX}{TensorFlow Extended}
	\acro{VM}{Virtual Machine}
	\acro{XML}{Extensible Markup Language}

	\end{acronym}

\begin{frontmatter}
\title{The Pipeline for the Continuous Development of Artificial Intelligence Models -\\Current State of Research and Practice}

\author[UIBKAddress]{Monika Steidl}
\ead{monika.steidl@uibk.ac.at}
\author[UIBKAddress,SwedenAddress]{Michael Felderer}
\ead{michael.felderer@uibk.ac.at}
\author[HagenbergAddress]{Rudolf Ramler}
\ead{Rudolf.Ramler@scch.at}


\address[UIBKAddress]{University of Innsbruck, 6020 Innsbruck, Austria}
\address[SwedenAddress]{Blekinge Institute of Technology, 371 79 Karlkskrona, Sweden}
\address[HagenbergAddress]{Software Competence Center Hagenberg GmbH (SCCH), 4232 Hagenberg, Austria}

\begin{abstract} 
Companies struggle to continuously develop and deploy \ac{AI} models to complex production systems due to \ac{AI} characteristics while assuring quality. To ease the development process, continuous pipelines for \ac{AI} have become an active research area where consolidated and in-depth analysis regarding the terminology, triggers, tasks, and challenges is required. \\ 
This paper includes a \acf{MLR} where we consolidated 151 relevant formal and informal sources. In addition, nine-semi structured interviews with participants from academia and industry verified and extended the obtained information. Based on these sources, this paper provides and compares terminologies for \ac{DevOps} and \ac{CI}/\ac{CD} for \ac{AI}, \acf{MLOps}, (end-to-end) lifecycle management, and \acf{CD4ML}. Furthermore, the paper provides an aggregated list of potential triggers for reiterating the pipeline, such as alert systems or schedules. In addition, this work uses a taxonomy creation strategy to present a consolidated pipeline comprising tasks regarding the continuous development of \ac{AI}. This pipeline consists of four stages: \textit{Data Handling}, \textit{Model Learning}, \textit{Software Development} and \textit{System Operations}. Moreover, we map challenges regarding pipeline implementation, adaption, and usage for the continuous development of \ac{AI} to these four stages.
\end{abstract}

\begin{keyword}
continuous development of AI \sep continuous (end-to-end) lifecycle pipeline for AI\sep MLOps\sep CI/CD for AI\sep DevOps for AI\sep multivocal literature review

\MSC[2022] 08-09\sep  99-00
\end{keyword}

\end{frontmatter}


\section{Introduction}
\label{chp:introduction}



An increase in available data and computing power, as well as improving algorithms, allow exploring the options of \ac{AI} in many different application fields. \ac{AI} and its subcategories, \ac{ML} and \ac{DL}, enable new intelligent products and services to achieve a specific goal \cite{Boucher.2020}. \\
To harness the power of \ac{AI}, it is necessary to deploy and integrate \ac{AI} models into production systems and to assure the quality of the resulting continuously evolving and self-adapting systems \cite{Fursin.2020, Stone.2016, Tao.2019, Pieters.2011}. However, quality assurance requires thorough attention to guarantee safe and reliable behaviour and increase the accountability and responsibility of the involved \ac{AI} systems \cite{Pieters.2011, Tao.2019, Hand.2020, Lenarduzzi.2020}. \ac{AI} characteristics such as the inherent non-determinism lead to a certain degree of uncertainty \cite{Stone.2016, Tao.2019, Pieters.2011}.\\
One possible solution for ensuring quality during the development of \ac{AI} are automated end-to-end \ac{CI}/\ac{CD} lifecycle pipelines \cite{Mishra.2020}. These pipelines are well established in traditional software development however need more research when adapting them to \ac{AI} models because these pipelines not only need to handle code but also data and the model itself in addition to a large system-level complexity \cite{fischer2020ai, Granlund.2021}. These pipelines focus on automating and monitoring all phases of system development, such as the integration, testing, and deployment, as well as the infrastructure management.\\
These pipelines for the continuous development of \ac{AI} are currently highly researched, where a synthesis of the current research provides an evidence-based foundation of the established work to avoid misconceptions, discover gaps in the knowledge field and assist research in exploring the phenomenon with further studies. Thus, the main goal of this paper is to \textbf{systematically identify relevant conceptual ideas, as well as synthesize and structure research in the area of pipelines for the continuous development of \ac{AI}}. Three research questions (RQ) have been derived from this overall goal, which we answer in this paper via a \acf{MLR} and follow-up interviews with practitioners from academia and industry.\\

\label{researchQuestions}
\begin{enumerate}
  \item Which terms are commonly used to describe a pipeline for continuous development of \ac{AI}? How do these terms differ in their specific meaning?
  \item Which tasks have to be handled by pipelines for the continuous development of \ac{AI}? What are possible triggers for starting the pipeline for continuous development of \ac{AI}?
  \item What are potential challenges when implementing, adapting, and using a pipeline for continuous development of \ac{AI}?
 \end{enumerate}

The remainder of the paper is structured as follows: Section \ref{background} presents necessary background knowledge on continuous software engineering and \ac{DevOps}. 
Furthermore, this section also provides an overview of related work with regard to the pipelines for the continuous development of \ac{AI} and describes the novel contribution provided by this paper.
Section \ref{chp:methodology} explains the information extraction and taxonomy creation methodologies. Section \ref{chp:results} introduces the terminology, triggers, taxonomy, and challenges. Section \ref{chp:discussion} applies \ac{TFX}, a lifecycle pipeline for \ac{AI}, to the proposed taxonomy. Section \ref{sec:threats} handles the threats to validity and Section \ref{chp:conclusion} concludes the paper and introduces future work.

\section{Background and Related Work}
\label{background}

\subsection{Continuous Software Engineering and DevOps}

This section broadly specifies the main background knowledge required for this paper. Firstly this section covers the general terms continuous software engineering, \ac{CI}/\ac{CD}, and \ac{DevOps} followed by a more detailed description of these terms in the context of \ac{AI}. For a more detailed description of the \ac{AI} related terms, please see Section \ref{sec:terminology}.\\

According to the established roadmap for continuous software engineering by Fitzgerald and Stol \cite{Fitzgerald.2017}, continuous software engineering describes the continuous development lifecycle, which includes continuous practices and concepts, such as \acf{CI}, \acf{CD}, continuous delivery and \acf{DevOps} \cite{Fitzgerald.2017}. \ac{CI} is a process that focuses on integrating code changes to the main software repository while automatically ensuring software quality \cite{Spieker.2019, Yasar.2020, informal_Pentreath.2019, Stahl.2014}. \ac{CD} describes the tasks after \ac{CI} and delivers or releases the new and tested software features to a staging or test environment \cite{gmeiner2015automated, Karamitsos.2020, Yasar.2020, informal?_Vadavalasa.2020}. Continuous Deployment requires that \ac{CD} already deployed the software to some environment other than production to ensure that the software can be continuously and automatically deployed to the production environment and to the actual users \cite{Karamitsos.2020, Yasar.2020, Fitzgerald.2017}. \ac{CI}/\ac{CD} for \ac{AI} are techniques to automate the deployment process for \ac{AI} models \cite{Borg.2021, Zhang.2020} (see Section \ref{subsec:CI/CDforML}.\\
\ac{DevOps} is a continuous software development approach that includes several principles and practices, such as \ac{CI}, \ac{CD}, and continuous deployment to manage a software system lifecycle. The term consists of \ac{Dev} and \ac{Ops}. \ac{Dev} uses agile methods, such as Scrum or Kanban, and allows a self-directed and self-organized software development with several teams \cite{Stirbu.2021, Yasar.2020, Kim.2016}. \ac{Ops} includes the tasks necessary to run an application, such as infrastructure management \cite{Yasar.2020}. \ac{DevOps} for \ac{AI} not only takes into consideration traditional software development but focuses on the added complexity of \ac{AI} development, such as data handling \cite{Rausch.2019} (see Section \ref{subsec:Devops}.\\
\ac{MLOps} expands \ac{DevOps} and takes into consideration the added complexity of developing \ac{ML} based applications \cite{informal_Google.2021} (see Section \ref{subsec:MLOps}.\\
The (end-to-end) lifecycle management describes the handling of specific tasks for the continuous development of AI, which starts with data collection and finishes with the deployment and monitoring of the \ac{AI} model \cite{AguilarMelgar.2021, Brumbaugh.2019, Zhou.2020} (see Section \ref{subsec:lifecycleManagement}).\\
\acf{CD4ML} is the technical implementation of \ac{MLOps} concept to automate the pipeline \cite{Makinen.2021, informal_Shtelma.2020} (see Section \ref{subsec:CD4ML}).

\subsection{Related Work} 
\label{sec:relatedWork}
Over the past years, a large number of publications focused on the topic of pipelines for the continuous development of \ac{AI} (e.g., \cite{Tamburri.2020, fischer2020ai, Renggli.2021, Makinen.2021, Makinen.2021b, Alnafessah.2021, Nguyen-Duc2020, Kolltveit2022, Nguyen-Duc2020, Amershi.2019, Martinez-Fernandez2022, Washizaki2019, Lewis2021}) as well as data handling \cite{Munappy2020, ManuelRodriguez.2020, Ereth}. In Table \ref{tab:relatedWork}, we list work that is most closely related to our study regarding the continuous development of \ac{AI} due to their similar methodology (\acf{SLR} and \acf{MLR}). The table describes the scope and research goal of related studies, and it indicates how the work maps to the three research questions addressed by our paper.

\renewcommand{\arraystretch}{1.2}
\begin{table}[hbt!]
\centering
\small
\begin{tabular}[t]{>{\raggedright\arraybackslash}p{2,3cm}>{\raggedright\arraybackslash}p{1,5cm}p{0,5cm}p{1cm}p{0,5cm}p{0,5cm}p{0,5cm}>{\raggedright\arraybackslash}p{4cm}}
\toprule
Reference & Research method & \# Papers & Review period & RQ1: terms & RQ2: tasks & RQ3: challenges & Differences in scope or goal\\
\midrule
\cite{Karamitsos.2020} Karamitsos et al. (2020) & \ac{SLR} & - & - &   & $\bullet$ &   & applied traditional \ac{DevOps} practices to \ac{AI}\\ 
\cite{John.2021} John et al. (2021) & \ac{MLR} & 29 & 1.2010-8.2020 &    & $\bullet$ & $\bullet$ & context: edge/cloud/hybrid architectures\\
\cite{John.2021b} John et al. (2021) & \ac{MLR} & 19 & 1.2015-3.2021 &   & $\bullet$ &   & maturity model based on \ac{MLOps}\\
\cite{Lwakatare.2020b} Lwakatare et al. (2020) &  \ac{MLR} \& interviews & 8 & - &   & $\bullet$ &   & how well \ac{CD} is applied to \ac{ML}-enabled systems\\
\cite{Figalist.2020} Figalist et al. (2020) & \ac{SLR} \& framework & - & - &   & $\bullet$ & $\bullet$ & context: ML-based software analytics/BI solutions\\ 
\cite{Lo2021} Lo et al. (2021) & \ac{SLR} incl. grey lit. & 231 & 1.2016-1.2020 &   &   &   & context: federated learning\\
\cite{Nascimento.2020} Nascimento et al. (2020) & \ac{SLR} & 55 & 1990-2019 &   &   & $\bullet$ & relationship between SE practice \& \ac{AI} development \\
\cite{Mboweni2022} Mboweni et al. (2022) & \ac{SLR} & 60 & 2015-2022 & $\bullet$ & && term \ac{MLOps} and main themes in literature\\
\cite{Fredriksson} Fredriksson et al. (2020) & \ac{SLR} & 43 & before 12.2019 &   &   &   & context: (semi-) automatic labelling of data types for \ac{ML}\\
\cite{Testi2022} Testi et al. (2022) & \ac{SLR} & - & 2015-2022 & & $\bullet$ & $\bullet$ & classify pipeline types, challenges for \ac{AI} development in pipeline\\
\cite{Kolltveit2022} Kolltveit et al. (2022) & \ac{SLR} & 24 & after 2015 & & $\bullet$ & $\bullet$ & operationalise \ac{AI}\\
\cite{Kreuzberger2022} Kreuzberger et al. (2022) & \ac{SLR} \& interviews & 27 & before 5.2021 & $\bullet$ & $\bullet$ & $\bullet$ & preprint only: principles and profession for realizing \ac{MLOps}  \\
\cite{informal_Lorenzoni.2021} Lorenzoni et al. (2021) & \ac{SLR} incl. grey lit. & 33 & 1.2010-6.2020 &   & $\bullet$ &   & preprint only; applicability of SE and \ac{ML} per author, no aggregated taxonomy of tasks\\ 
\cite{informal_Xie.2021} Xie et al. (2021) & systematic mapping study & 405 & before 7.2020 &   &   &   & preprint only; mapping study of \ac{AI} model lifecycle management (no focus on tasks) \\
\midrule
Our study & \ac{MLR} \& interviews & 151 & 2010-2021 & $\bullet$ & $\bullet$ & $\bullet$ & \ac{AI} pipelines: definition of terms, taxonomy of tasks, challenges\\
\bottomrule
\end{tabular}
\caption{Overview of related work ($\bullet$ indicates full or partial coverage of targeted RQs)}
\label{tab:relatedWork}
\end{table}
\renewcommand{\arraystretch}{1}

\FloatBarrier

The following section provides an extensive analysis of the mentioned related work from Table \ref{tab:relatedWork}.\\
Firstly, several related studies \cite{Karamitsos.2020, John.2021, John.2021b, Lwakatare.2020b, Figalist.2020, Fredriksson, Kolltveit2022, Kreuzberger2022, informal_Lorenzoni.2021} are based on a limited amount of identified primary sources, not covering relevant insights from the wide range of existing literature. Our study is based on a comprehensive analysis including over 150 papers. 
\\
Secondly, some literature reviews in related work focus on a specific application context such as edge/cloud/hybrid architectures \cite{John.2021}, \ac{ML}-based software analytics and business intelligence applications \cite{Figalist.2020}, or federated learning \cite{Lo2021}. In contrast, our study covers the full scope of \ac{AI} models, independently of a specific application context.
\\
Thirdly, a range of different research questions are investigated in related studies, not or only partially related to the definition of terms (RQ1), pipelines for the continuous development of \ac{AI} and triggers for starting the pipeline (RQ2), and pipeline-related challenges (RQ3).
\begin{enumerate}
\item \textbf{Definition of terms}: \ac{SLR} or \ac{MLR} based papers in the identified related work do not elaborate on the definition of terms for continuous development of \ac{AI}. \textit{Mboweni et al.}~\cite{Mboweni2022} and \textit{Kreuzberger}~\cite{Kreuzberger2022} provide a foundation-based definition on \ac{MLOps}. Definitions for related terms (e.g., \ac{CI}/\ac{CD} for \ac{AI}) are not considered. 

\item \textbf{Pipelines for the continuous development of \ac{AI}}: Related studies target various different approaches and research goals. For instance, \textit{Karamitsos et al.}~\cite{Karamitsos.2020} base their \ac{CI}/\ac{CD} pipeline for \ac{AI} on a literature review focusing on "traditional" \ac{DevOps} principles, not covering \ac{AI} specific tasks due to the non-existence in \ac{DevOps} pipelines. Generally, less emphasis is placed on tasks necessary to develop \ac{AI} continuously. \textit{John et al.}~\cite{John.2021b} proposes a \ac{MLOps} maturity model consisting of tasks for data handling, development and release of the \ac{ML} model. \textit{Lwakatare et al.}~\cite{Lwakatare.2020b} executed a \ac{MLR} to identify how well \ac{CD} is applied to \ac{ML}-enabled systems and proposed levels of automation, where the first level indicates the manual process and the fifth level is the fully automated and integrated process where \ac{CD} is incorporated into the \ac{ML} workflow process \cite{Lwakatare.2020b}. \textit{Fredriksson et al.}~\cite{Fredriksson}, for example, also executed a \ac{SLR} but only cover approaches to label different data types to be used for supervised training. \textit{Testi et al.}~\cite{Testi2022} summarize different types of \ac{MLOps} pipelines, such as \ac{ML}-based software systems, \ac{ML} use case applications, \ac{ML} automated framework where tasks of this automated framework are briefly summarized. \textit{Kolltveit et al.}~\cite{Kolltveit2022} do not consider all required tasks for generating an \ac{AI} model but focus on operationalising the model via packaging, integration, deployment, serving, inference and monitoring and logging. \textit{Kreuzberger et al.}~\cite{Kreuzberger2022} cover principles within technical components (e.g., reproducibility achieved via feature store) and required professions to build the pipeline. \textit{Lorenzoni et al.}~\cite{informal_Lorenzoni.2021} identified which software engineering processes and practices can be applied to solve issues arising during the development of \ac{ML} models. However, the paper does not contain an in-depth analysis of the \ac{CI}/\ac{CD} phases, and how developers implement them. Moreover, their paper does not consider continuous execution of such phases, as described in this paper. \textit{Xie et al.}~\cite{informal_Xie.2021} executed a systematic mapping study to identify demographic data, such as when and where the papers were published, which research methods were applied, and which subtopics were covered in the literature. However, they did not focus on the continuous development tasks but used search terms focusing on characteristics of the lifecycle, such as traceability, reproducibility, guidelines, and transparency.\\\
As indicated, DataOps is also addressed in related work by \textit{Rodriguez et al.}~\cite{ManuelRodriguez.2020}, \textit{Munappy et al.}~\cite{Munappy2020}, and \textit{Ereth}~\cite{Ereth} with the focus on the first step of the pipeline for the continuous development of \ac{AI}, namely data handling. With this paper, we expand their work by investigating all necessary tasks of the entire pipeline including the deployed and monitored of an \ac{AI} model.


\item \textbf{Triggers}: No information regarding triggers of the pipeline for the continuous development is covered by the related \ac{SLR}s and \ac{MLR}s.

\item \textbf{Pipeline related challenges}: Challenges covered by related work mostly focus on the development of \ac{AI} in general, but not on the pipeline itself. For instance, \textit{Nascimento et al.}~\cite{Nascimento.2020} illustrate the relationship and dependencies between software engineering practices based on the SWEBOK knowledge areas and respective challenges regarding the development of \ac{AI} systems. \textit{Kreuzberger et al.}~\cite{Kreuzberger2022} focus on organisational, \ac{ML} system, and only very superficially on operational challenges for adopting \ac{MLOps}, which is the core of this paper's challenges. \textit{Figalist et al.}~\cite{Figalist.2020} identify challenges during prototyping, deployment and update, but they specifically focus on \ac{AI} models for software analytics and business intelligence. \textit{Testi et al.}~\cite{Testi2022} illustrate \ac{AI} specific challenges which do not focus on the implementation of the continuous pipeline, such as data labelling. \textit{Kolltveit et al.}~\cite{Kolltveit2022} cover exclusively challenges regarding operationalising the \ac{AI} model.\\
Related work, which does not follow the same methodology as our study also focuses on challenges occurring throughout the pipeline for the continuous development of \ac{AI} systems. These studies, however, identify general challenges of individual tasks during the development process of \ac{AI} applications. For instance, \textit{Paleyes et al.}~\cite{informal_Paleyes.2020} identify data collection as a challenge, covering issues regarding related storage location and understanding the data set's structure. \textit{Baier et al.}~\cite{Baier.2019} differentiate challenges occurring during the pre-deployment (e.g., data structure, data quality, and governance) and deployment (e.g., detecting and handling data drifts) as well as non-technical challenges (e.g., expectation management, trust, transparency). \textit{Lewis et al.}~\cite{Lewis2021} only briefly mention general challenges regarding \ac{ML} system development, summarized to data management, modelling and operationalisation, with the aim to evaluate how well available tools are able cope with these challenges.
\end{enumerate}

\section{Methodology}
\label{chp:methodology}
\begin{figure}[!ht]
	\centering
	\includegraphics[width=\linewidth]{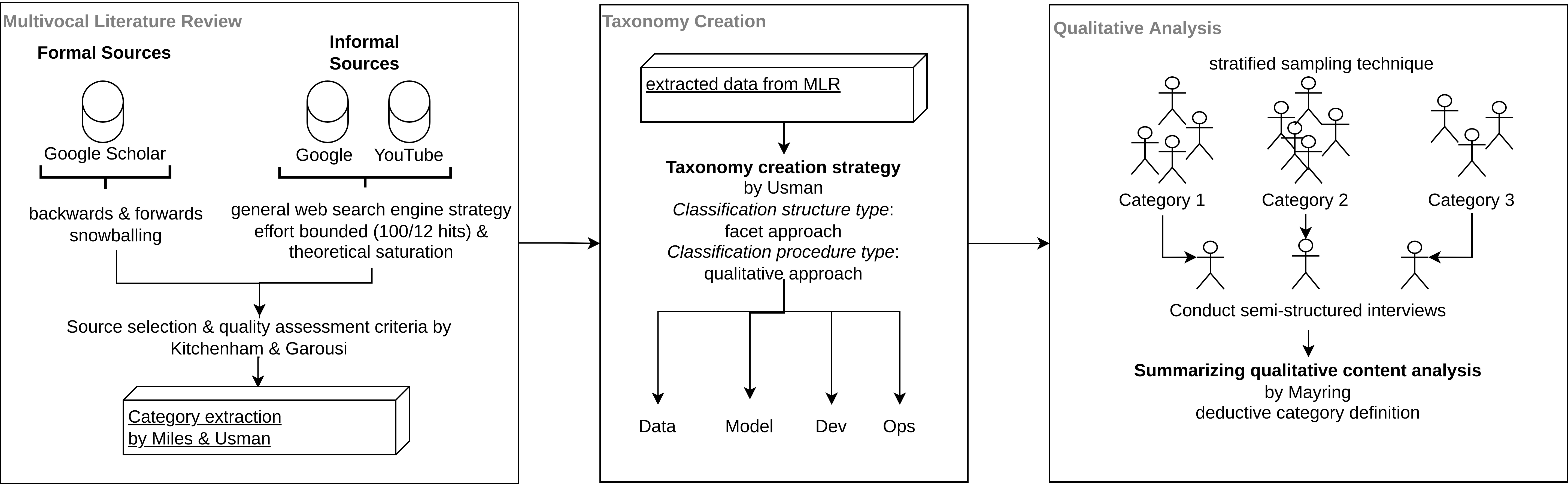}
	\caption[Study design for \ac{MLR}, taxonomy development method and deductive category definition]{Study design for \ac{MLR} based on \cite{Garousi.2019}, taxonomy development method based on \cite{Usman.2017}, and deductive category definition based on \cite{Mayring.2015}}.
	\label{fig:MLRSurvey}
\end{figure}

In the following, this section covers the three employed research methods (\ac{MLR}, taxonomy creation strategy and qualitative analysis) to derive the main research contributions. Figure \ref{fig:MLRSurvey} illustrates a general overview of the research design consisting of a \ac{MLR} proposed by Garousi et al. \cite{Garousi.2019} to identify existing literature, taxonomy development method based on Usman et al. \cite{Usman.2017} to map the identified aspects in literature to a taxonomy, and the interviews' deductive category definition based on Mayring \cite{Mayring.2015}. These research methods are further explained in the following sections.

\subsection{MLR}
A \ac{MLR} was executed to provide a thorough overview and aggregated evidence of important perspectives of pipelines for the continuous development of \ac{AI}. This allows not only to include published literature but also allows to include grey literature. Grey literature is essential because using a lifecycle pipeline for \ac{AI} is an emerging research topic in software engineering where formal literature has not been sufficiently published yet \cite{Garousi.2019}. The following approach is based on the paper `Guidelines for including grey literature and conducting multivocal literature reviews in software engineering` provided by Garousi et al. \cite{Garousi.2019}.

Table \ref{table:selectionCriteria} depicts the general \textbf{selection criteria} that apply to both selection processes of formal as well as informal sources. 

\begin{table}[h!]
\centering
\begin{tabular}[t]{p{2cm}p{4.5cm}p{4.5cm}}
\toprule
 & Inclusion Criteria & Exclusion Criteria \\
\midrule
\textit{Year of publication} & 2010-2021 & before 2010 \\
\textit{Language} & English & Any other language \\
\textit{Accessibility} & Full text needs to be accessible & Parts of source available \\
\textit{Relevance} & Relevant information to answer the main research questions (e.g. Does the source focus on pipelines for the continuous development of \ac{AI}) & Sources that focus on using \ac{AI} to implement \ac{DevOps} such as \ac{AIOps} or which focus only on team related processes (e.g. team collaboration) \\
\bottomrule
\end{tabular}
\caption{General selection criteria for formal and informal sources}
\label{table:selectionCriteria}
\end{table}

Table \ref{table:selectionCriteriaformal} depicts specific selection criteria for \textbf{formal sources}. Based on these criteria we derived 37 formal sources as the start data set. Afterwards, we followed Wohlin's proposed backward and forward snowballing procedure \cite{Wohlin.2014}. We derived the final data set that comprises 79 papers between the 15th of April and the 30th of May 2021. When citing formal sources, this paper uses the prefix F in combination with a number.\\
In order to retrieve \textbf{formal/scientific sources}, we executed an initial exploratory search with several search terms (see \ref{searchterms} to collect a start data set \cite{Garousi.2019}. We followed the guidelines proposed by Kitchenham \cite{Kitchenham.2007} and Wohlin and Jalali \cite{Wohlin.2014, Jalali.2012}. We used Google Scholar as the search engine to retrieve an unbiased start data set. According to Yasin et al. \cite{Yasin.2020} results from Google Scholar in combination with grey literature extracted from Google sufficiently extracts necessary sources similar to searches with other databases, such as ScienceDirect, IEEE, ACM digital library and Springer Link.\\
\\

\begin{table}[h!]
\centering
\begin{tabular}[t]{p{2cm}p{4.5cm}p{4.5cm}}
\toprule
 & Inclusion Criteria & Exclusion Criteria \\
\midrule

\textit{Quality criteria} & Primary peer-reviewed sources & No review process, secondary studies \\
Search strategy & Google Scholar & Other databases \\
\textit{Stopping criteria} & First 100 search hits & Search hits after \\
\bottomrule
\end{tabular}
\caption{Selection criteria for formal sources}
\label{table:selectionCriteriaformal}
\end{table}

Table \ref{table:selectionCriteriaInformal} depicts the specific selection criteria for \textbf{informal sources}. Based on these criteria we derived informal sources between the 31st of May and the 26th of June 2021. When citing informal sources, the prefix
I is used in combination with a number.\\
We adopt \textit{quality assessment criteria} to select the sources based on their relevance and to check whether the informal literature search results are valid and free of bias \cite{Garousi.2019}.
We derive the sources based on the proposed \textit{general web search engine} strategy where we use conventional web search engines \cite{Garousi.2019}. Regarding the stopping criteria, we selected two strategies. Firstly, \textit{effort bounded} looks at a predefined number of search engine hits. Secondly, \textit{theoretical saturation} identifies whether no new concepts emerge from additional search results. \textit{Theoretical saturation} was not achieved in seven out of twenty cases. \\

\begin{table}[h!]
\centering
\begin{tabular}[t]{p{2cm}p{4.5cm}p{4.5cm}}
\toprule
 & Inclusion Criteria & Exclusion Criteria \\
\midrule
\textit{Quality criteria} & Fulfill Garousi et al.’s \cite{Garousi.2019} quality assessment criteria & deviation of Garousi et al.’s \cite{Garousi.2019} quality assessment criteria \\
Search strategy & General web search engine (Google, YouTube) & Specialized databases and websites (e.g. Stackoverflow), contacting individuals directly \\
\textit{Stopping criteria} & Effort bounded: Google (100 search hits), YouTube (12 search hits) & Search hits after effort bounded and theoretical saturation are fulfilled \\
& Theoretical saturation required, otherwise additional hits searched: Google (50 search hits), YouTube (20 search hits) \\

\bottomrule
\end{tabular}
\caption{Selection criteria for informal sources}
\label{table:selectionCriteriaInformal}
\end{table}

\begin{figure}[!ht]
	\centering
	\includegraphics[width=\linewidth]{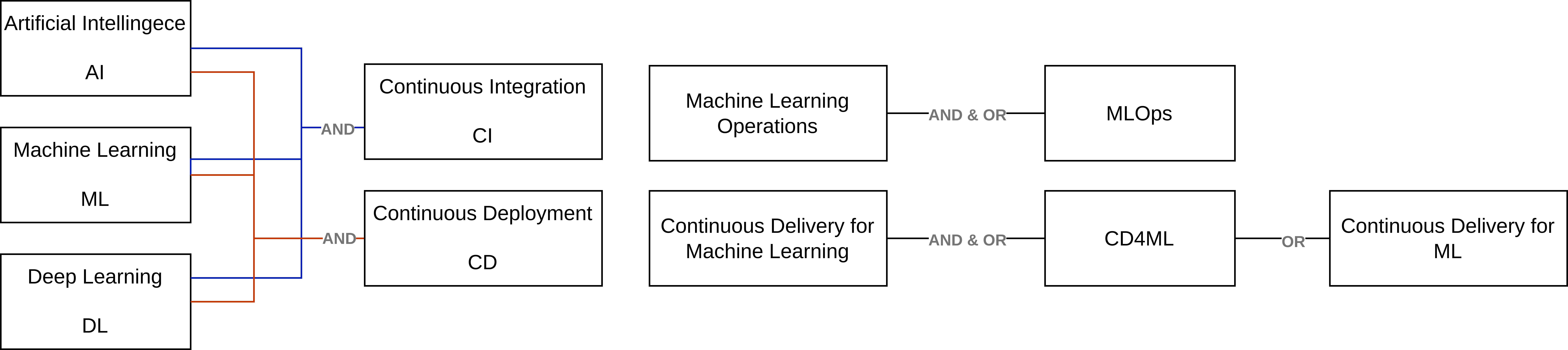}
	\caption[Search terms used for the \ac{MLR}]{Search terms used for the \acf{MLR}}
	\label{fig:searchTerm}
\end{figure}

\label{searchterms}
Figure \ref{fig:searchTerm} illustrates the used ten \textit{search terms} which were based on the established terms of continuous software engineering \cite{Fitzgerald.2017}. Additional search strings were extracted during an informal pre-search. Example for the used search strings are \texttt{Artificial Intelligence AI AND Continuous Integration CI}, \texttt{Machine Learning Operations AND MLOps}, and \texttt{Machine Learning\\ Operations OR MLOps}. \\

The \ac{MLR} identified 151 relevant sources, out of which 79 papers (approximately 53\%) were formal sources and 72 informal sources. The extraction process and the retrieved formal and informal literature were documented in a systematic map which is available online \cite{steidl_monika_2022_5902776}.\\

To \textbf{extract} the necessary \textbf{categories} from the literature, a we executed a descriptive qualitative synthesis. Kitchenham et al. \cite{Kitchenham.2007} require to document the extracted information in a tabulated and consistent manner based on the previously defined research questions. For further information on the tabulation of the extracted information, please refer to \cite{steidl_monika_2022_5902776}. We derive further subcategories of the research questions as suggested by Stol et al. \cite{Stol2016} and create the taxonomy as proposed by Usman et al. \cite{Usman.2017}. We adopted Stol et al.'s \cite{Stol2016} coding strategy, namely open and axial coding to break down, examine, compare, conceptualize, and categorize information. We base the categories on a previous pilot study. These categories were closely related to the terms used in continuous software engineering, hence commonly accepted within the field \cite{Usman.2017}. Because the predefined set of categories did not cover all tasks handled with the pipeline for the continuous development of \ac{AI}, we added additional categories via an iterative process.

After selecting the source and extracting necessary categories, we executed a \textbf{test-retest process} proposed by \cite{Kitchenham.2007} to evaluate the rater's data extraction consistency. Further information is provided in Section \ref{sec:threats}.

\subsection{Taxonomy Creation}
\label{subsec:TaxonomyCreation}
We categorize the extracted data via a taxonomy based on the revised taxonomy creation strategy proposed by Usman et al.'s \cite{Usman.2017}. Firstly, we defined the units of the classes/categories which are based on \ac{DevOps} phases because they are commonly accepted within the field. We add extracted information to the respective class/category via a qualitative approach \cite{Usman.2017}. For the classification structure type, we use a a facet approach because research on pipelines for the continuous development of \ac{AI} applications is still a new and evolving field. The facet approach allows us to easily adapt the taxonomy if further research is done on this topic. The identified facets comprise the stages \textit{Data Handling}, \textit{Model Learning}, \textit{Software Development} and \textit{System Operations}.\\

\subsection{Qualitative Analysis}
\label{subsec:qualitativeAnalysis}
We check via a qualitative approach if the derived information from the literature is correct and comprehensive enough to provide a thorough depiction of existing knowledge. For this, we conducted interviews because they allowed us to explore and understand individual experiences from a sample by outlining the complexity and diversity of the observed environment \cite{Miles.2014}. To select the interview participants, we adopt a stratified sampling technique \cite{Robinson.2014} with the three groups illustrated in Table \ref{table:categories}.

\begin{table}
\begin{tabularx}{\linewidth}{XXXX}
\toprule
& Category 1 & Category 2 & Category 3\\ \midrule
\textit{Area of Research}  &      Academia           & \begin{tabular}[c]{@{}l@{}} Industry \end{tabular} &     \begin{tabular}[c]{@{}l@{}}Industry\\(Start-ups)\end{tabular} \\\\
\textit{Experience with \ac{AI}}  &      Yes & Yes & Yes \\\\
\begin{tabular}[c]{@{}l@{}}\textit{Experience with}\\\textit{lifecycle pipelines}\end{tabular} & Yes & Yes & \begin{tabular}[c]{@{}l@{}}Should know\\concept of\\continuous \\lifecycle pipeline\end{tabular} \\\\
\textit{General experience} & \begin{tabular}[c]{@{}l@{}}Published work\\on lifecycle\\management for \ac{AI}\end{tabular}  & \begin{tabular}[c]{@{}l@{}} Extensive \& regular\\deployment of \ac{AI}\\developers\\(no data scientists)\end{tabular} & \begin{tabular}[c]{@{}l@{}}\ac{AI} development\\and deployment\end{tabular} \\
\end{tabularx}
\caption{Defined categories for the stratified sampling}
\label{table:categories}
\end{table}

Based on the selection criteria, we identified nine participants. Table \ref{table:participantDescription} provides an overview of the involved participants.

\begin{table}[ht!]
\centering
\begin{tabular}[t]{p{1cm}p{4cm}p{1cm}p{1,7cm}p{1,3cm}p{1cm}}
\toprule
Partici-pant & Knowledge about pipelines for the continuous development of \ac{AI} & Expe-rience & Industry & Region & Cate-gory\\
\midrule
A  &      Team lead for AI initiative un- \& supervised algorithms &  5.5 years       &     Social Media & America & 2 - Industry  \\\\
T  & Research projects: continuous software engineering practices for traditional and \ac{AI} software      &    10 years   & Academia & Finland & 1 - Aca-demia \\\\
Z & Sales Engineer for \ac{MLOps} platform, \ac{AI} development & 1 year & \ac{MLOps} software & America & 2 - Industry\\ \\
P & Team Lead \ac{AI} application engineering & 2 years & Text analysis & Austria & 3 - Industry (Start-ups)\\ \\    
R & Technical Team Lead for \ac{ML} deliveries (e.g. sentiment analysis, speech assistant)  & 3.5 years  & Automotive & Germany & 2 - Industry\\\\
B & Research (Area manager for services and solutions): \ac{AI} innovations \& image processing  & 5.5 years  & Research/ Consultancy & Austria & 2 - Industry \\ \\
C & Implementation of pipeline for the continuous development of \ac{AI} Experience in Kubernetes \& \Ac{ML}  & 4 years  & Modeld.io: \ac{MLOps} pipeline & America & 3 - Industry (Start-ups)\\ \\  
V & Research (Senior research project manager): pipeline for the continuous development of \ac{AI} with \ac{TFX} & 3 years  & Research/ Consultancy & Austria & 2 - Industry\\ \\
D & Regulatory compliance in \ac{MLOps} \& Certification body for \ac{AI} & 2 years  & Academia (Healthcare) & Finland & 1 - Aca-demia\\ \\ 
\bottomrule
\end{tabular}
\caption{Overview and background of interviewed participants}
\label{table:participantDescription}
\end{table}

The semi-structured interviews included introductory questions, questions about different definitions of pipelines for the continuous development of \ac{AI}, tasks handled via these pipelines, and an evaluation of the proposed taxonomy as well as challenges when implementing, adapting, and using such a pipeline. We conducted the interviews between the 30th of July and the 10th of September 2021 with an average duration of 49 minutes. We analysed the interviews according to Mayring \cite{Mayring.2015} with the \textit{summarizing qualitative content analysis}. We adopted the \textit{deductive category definition} to extract information regarding the pipeline. In addition, we categorized the information on theoretically derived aspects from the \ac{MLR} \cite{Mayring.2015}. Thus, we derive the coding agenda from the identified stages and tasks from the previously mentioned taxonomy creation. By doing so we unambiguously assign the participants' statements to the identified categories \cite{Mayring.2015}.
\section{Results}
\label{chp:results}
This section presents the information obtained from the literature and interviews. Firstly, the different terminologies are elaborated, followed by identified triggers to start/restart the pipeline. The following section elaborates on the created taxonomy which describes the pipeline for the continuous development of \ac{AI} and included tasks. The final subsection explores challenges regarding the implementation, adaption and usage of pipelines for the continuous development of \ac{AI}.

\subsection{Terminologies}
\label{sec:terminology}
\begin{figure}[!ht]
	\centering
	\includegraphics[width=\linewidth]{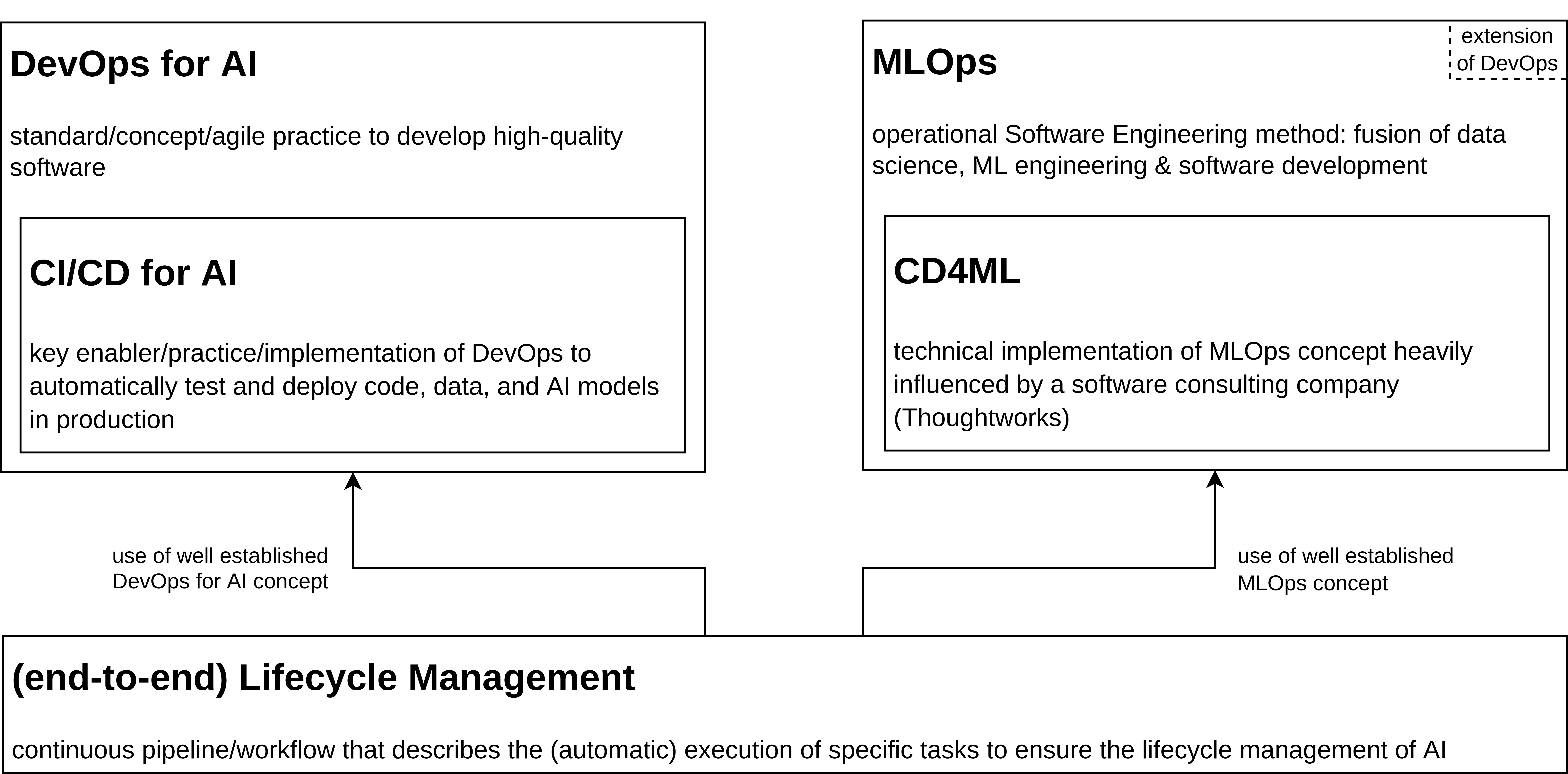}
	\caption{Summary of terms - \ac{DevOps} for \ac{AI}, \ac{CI}/\ac{CD} for \ac{AI}, \ac{MLOps}, (end-to-end) lifecycle management, and \ac{CD4ML}}
	\label{fig:terminology}
\end{figure}

The following subsections describes several terms, their main characteristics and differences, including (1) \ac{DevOps} for \ac{AI}, (2) \ac{CI}/\ac{CD} for \ac{AI}, (3) \ac{MLOps}, (4) (End-to-End) Lifecycle Management, (5) \ac{CD4ML}. Figure \ref{fig:terminology} illustrates the main characteristics of the terms. All terms share a common understanding and describe the automation for the continuous development and improvement of \ac{AI} models via pipelines. The interviewees indicated that the differences between the terms are unknown in practice. According to three of the interviewed participants, the terms describe the process of adapting the standard software lifecycle to \ac{AI} modelling in order to minimize the time between iterations and to ease the whole process of continuously developing and deploying \ac{AI} models.\\

\subsubsection{DevOps for AI}
\label{subsec:Devops}
Some authors use the term \ac{DevOps} for AI to describe \textit{'[...] methods for managing the software lifecycle'} \cite{Rausch.2019} as stated by Rausch et al. It is seen as a standard for modern software development to ensure higher data, as well as code quality \cite{Rausch.2019, informal_Breuel.2020}. \ac{DevOps} for \ac{AI} is a concept and agile practice to reduce time and resources between deployment iteration cycles \cite{Rausch.2019, Karlas.2020, informal_Breuel.2020}.

\subsubsection{CI/CD for AI}
\label{subsec:CI/CDforML}
\ac{CI}/\ac{CD} are key enablers or techniques for \ac{DevOps} to stabilize, optimize and automate the deployment process of \ac{AI} models \cite{Borg.2021, Zhang.2020, informal_Baroni.2018}. According to Karlas et al. \ac{CI}/\ac{CD} supports the \textit{'[...] deployment to the infrastructure used to serve models in production'} \cite{Karlas.2020}. Thus, \ac{CI}/\ac{CD} takes into consideration not only validating and testing code and components, but also handles the (semi-)automatic and iterative validation and testing of data, data schemas, and models \cite{informal_Google.2021, Karlas.2020}. 

\subsubsection{MLOps}
\label{subsec:MLOps}
According to Google Cloud \ac{MLOps} \textit{'[...] appl[ies] \ac{DevOps} principles to \ac{ML} systems'} \cite{informal_Google.2021} and participant Z and R confirm that \ac{MLOps} is an extension of \ac{DevOps}. According to Raj et al. it is an \textit{'[...] emerging method to fuse machine learning engineering with software development'} \cite{Raj.2021}. This statement is underlined by other sources as well
\cite{Raj.2021, Martel.2021, Sangiovanni.2020, Makinen.2021, Zhou.2020, MartinezFernandez.2021, Renggli.2021}. Breuel defines it as following: \textit{'\ac{MLOps} is a set of practices that combines \ac{ML}, \ac{DevOps} and Data Engineering, which aims to deploy and maintain \ac{ML} systems in production reliably and efficiently'} \cite{informal_Breuel.2020}.\\
The key difference between other terms is that \ac{MLOps} strongly takes into consideration the company's culture and illustrates how cross-functional teams, such as data analysts, system operators, as well as data and software engineers collaborate via a harmonized process \cite{Sangiovanni.2020, Poloskei.2020, Junsung.2019, MartinezFernandez.2021, Xu.2020, Stirbu.2021, informal_Tandon.2021, informal_keating.2020}. This statement was confirmed by participant T and R.\\
Similar to \ac{DevOps} for \ac{AI}, \ac{MLOps} helps with the continuous, quick, seamless and reliable deployment of multiple \ac{AI} versions which are deployed in a heterogeneous and distributed environment via infrastructure and tools \cite{Raj.2021, Bourgais.2021, MartinezFernandez.2021, Fursin.2020}. A new practice, called \ac{CT} is introduced that according to Google Cloud \textit{'[...] is concerned with automatically retraining and serving the models'} \cite{informal_Google.2021}. Therefore, \ac{CT} uses collected feedback and production data \cite{informal_Google.2021, Karlas.2020, informal_Google.2021, informal_GoogleTFX.2021, informal_Mulkens.2020}.\\
The \ac{AI} model quality is strongly dependent on the used data sets and the model is only a small part of the entire software system \cite{Yasar.2020, Yasar.2020b, Makinen.2021, informal_Haviv.2020, informal_Patel.2019, informal_Ammanath.2021, Renggli.2021}. This statement was underlined by participant Z.\\
According to the Google Cloud documentation \cite{informal_Google.2021}, \ac{MLOps} can be divided into three different levels of maturity depending on the degree of automation. Another definition by Microsoft was identified during the interviews where five levels of technical implementation of \ac{MLOps} are defined\footnote{Microsoft's maturity levels for \Ac{MLOps}: \url{https://docs.microsoft.com/en-us/azure/architecture/example-scenario/mlops/mlops-maturity-model}, accessed 16.12.2021}.

\subsubsection{(End-to-End) Lifecycle Management}
\label{subsec:lifecycleManagement}
This term includes the word \textbf{management}, which describes the handling of specific tasks of the continuous development of \ac{AI} \cite{Vartak.2016, informal_Katsiapis.2020}. Essential management tasks start with data collection and finish with \ac{AI} model deployment and monitoring in production \cite{AguilarMelgar.2021, Brumbaugh.2019, Zhou.2020, Chard.2019, Miao.2017c}, as verified by participant T. Vartak et al. further specifies tasks included in the model management such as  \textit{'[...] tracking, storing and indexing large numbers of machine learning models so they may subsequently be shared, queried and analyzed'} \cite{Vartak.2016}.\\
Based on these tasks, the \textbf{(end-to-end) lifecycle} for \ac{AI} models describes a well-fitted pipeline that should achieve the best possible quality and stability of the \ac{AI} components via several iterations until the model cannot be improved any further \cite{Karlas.2020, Brumbaugh.2019, Miao.2017, Miao.2017b}. During these iterations, several data sets, artifacts, models and application configurations are created, which need to be managed, searched, shared and analysed \cite{Vartak.2016, Zhou.2020}. For example, it is crucial for Brumbaugh et al. to \textit{'[...] have correct values for the features that correspond to the timestamp of the labels'} \cite{Brumbaugh.2019}. \\ 
Several authors use the term continuous pipeline or workflow to describe the automatic execution and reiteration of tasks to ensure the lifecycle management of \ac{AI} \cite{Barrak.2021, Miao.2017b, Zhou.2020, Lwakatare.2020, Baylor.2019}. According to Barrak et al. a \textit{'[...] pipeline of tools [...] automate[s] the collection, preprocessing, cleaning and labelling of data.'} \cite{Barrak.2021}

The term end-to-end lifecycle management benefits from the idea of automation for the whole model lifecycle. The term uses well-established concepts from software development to cope with many model iterations, such as \ac{DevOps} in combination with \ac{CI}/\ac{CD} \cite{Zhou.2020, Bachinger.2020, informal_Aronchick.2020} and \ac{MLOps} \cite{informal_Patel.2019}. However, in contrast to \ac{MLOps}, lifecycle management does not focus on the interpersonal collaboration between different teams \cite{Miao.2017c}.\\

\subsubsection{CD4ML}
\label{subsec:CD4ML}
\ac{CD4ML} is a technical implementation of the \ac{MLOps} concept to automate the pipeline for the continuous development of \ac{AI} \cite{Makinen.2021, Granlund.2021, informal_Sato.2019, informal_SatoMartinFowler.2019, informal_Shtelma.2020}. Therefore, \ac{CD} principles are used to span the \ac{AI} lifecycle management and apply them to \ac{AI} applications \cite{Makinen.2021, informal_Gorcenski.2019, informal_Windheuser.2020}. Participant T as well as the extracted information from the literature identify that this term is proposed, promoted and heavily influenced by Thoughtworks \cite{Makinen.2021, Stirbu.2021, informal_Windheuser}, which defines \ac{CD4ML} as \textit{'[...] a software engineering approach in which a cross-functional team produces machine learning applications based on code, data, and models in small and safe increments, that can be reproduced and reliably released at any time in short adaptation cycles'} \cite{informal_Sato.2019}.

\subsection{Triggers}
\label{sec:triggers}
The following section discusses four trigger types, including (1) feedback and alert systems, (2) orchestration service and schedule, (3) repository, and (4) other triggers.
\ac{AI} models need to be iteratively adapted and retrained to provide reliable quality in production over a long period of time \cite{Baylor.2017, Baylor.2019, informal_Baroni.2018, informal_Arnold.2020}. Therefore, according to Moesta et al. \cite{informal_Moesta.2020} and two participants (R and Z), context-specific triggers depending on the \ac{AI} model, business requirements and retraining strategies exist that start or restart the pipeline. For example, triggers may take into consideration the optimal threshold where the benefits obtained by an updated (i.e., retrained) model outweigh the effort involved in the retraining~\cite{Baylor.2017, informal_Ettun.2019, Kronberger.2020, Schelter.2018, informal_Rausch.2020}. Participants R and D indicated that it is a trial and error process to minimize resource consumption where several different team members identify the appropriate approach. Thus, triggers combining different approaches may also be feasible \cite{Rausch.2019, Raj.2021}.

\subsubsection{Feedback and Alert Systems}
\label{subsec:FeedbackAllertTrigger}
Collected feedback during runtime or alerts may be used to trigger the pipeline \cite{Derakhshan.2019}. Three interviewees identified data events as a potential trigger, whereas information extracted from literature also covers data and model changes \cite{Rausch.2019, informal_Visengeriyeva.2021, informal_Schruhl.2020}.\\

A monitoring system monitors and collected \textbf{data} from production to trigger the pipeline in case of irregular data events such as data updates and data drifts \cite{Breck.2019, Baylor.2019}. Data drifts occur when the distribution within the data set changes \cite{informal_Gorcenski.2019, Vuppalapati.2020, Garcia.2018, informal_Erb.2019}. This occurs when data varies due to seasonal changes, or any other insertion, deletion or update of data values \cite{informal_Baroni.2018, Martel.2021, Liu.2019, Baylor.2017, Zhou.2020}. The interview participants highlighted the deletion of data. For example, participant T mentioned that due to privacy restrictions and the data regulation requirements in Europe, users have the right that their associated data is forgotten. Thus, according to participant T ' [...] it is only fair that the deleted data is no longer used in the \ac{ML} model'. Data updates may also happen if the shape of the data, such as table or constraint definitions, may change due to schema updates based on software updates, requirement changes or migrations \cite{informal_Gorcenski.2019}.\\
According to participants T, R and D, data updates should improve the model. To avoid triggering the pipeline continuously, triggers may occur periodically or when a specific threshold is attained \cite{Amershi.2019, Polyzotis.2017, Breck.2019}. Similar to the results obtained from the literature, it is ambiguous for participant D what the appropriate amount of new data to change the outcome of a model is. According to participant R, the changes need to be extensive enough to significantly impact the model. Not mentioned in the extracted literature's information, however, indicated by participant R, is that event streaming platforms, such as Apache Kafka or other event hubs, may be used to semi-automate the triggering process.\\

\textbf{Model} updates may be triggered due to the deterioration of the model's performance, and scores in production below a specific threshold, also called model or concept drift \cite{Martel.2021, Zhou.2020, Janardhanan.2020, LopezGarcia.2020, Renggli.2019, informal_Xin.2021}. It occurs when the problem the model was designed to solve changes, and this problem needs to be reformulated \cite{informal_Visengeriyeva.2021, informal_Srinivasan.2021, informal_Rosenbaum.2020}. Technical performance scores indicating a trigger are throughput, latency, and the utilization of a \ac{GPU} \cite{Zhang.2020, informal_Baroni.2018}. Two participants (Z and C) use these performance metrics as triggers. 

\subsubsection{Orchestration Service and Schedule}
\label{subsec:OrchestrationTrigger}
Additional triggers are an orchestration service or a scheduled time \cite{Rausch.2019, Brumbaugh.2019, Raj.2020}. For example, one participant triggered their pipeline once a week. On the one hand, one may argue that fixed schedules hinder the pipeline to be reactive enough or needless pipeline executions are triggered \cite{Baylor.2019}. On the other hand, schedules help to optimize the retraining frequency, allocation of computing resources and execution order of pipeline jobs, which is especially important if edge and cloud resources are involved \cite{Boag.2017, informal_Google.2021, LopezGarcia.2020, Rausch.2019b}.\\

\subsubsection{Repository}
\label{subsec:RepositoryTrigger}
Repository updates are used as traditional triggers to guarantee that the latest changes are tested and available to the users. For example, pull requests to the repository as a commit or merge requests identify changes to the data, model or code \cite{informal_Aronchick.2020, informal_Shtelma.2020, informal_Windheuser, informal_Erb.2019, informal_Patel.2020, informal_Sierra.2018}. When using this approach, participant T indicated that the data set and the appropriate code should be in the same system and under the same source control.

\subsubsection{Other Triggers}
Although manual triggers are sparsely elaborated in the collected literature \cite{Lwakatare.2020b, Raj.2020, informal_Liu.2020, informal_Aronchick.2020}, three out of nine participants use triggers that involve human interaction to identify if sufficient data is available. One of the reasons is that the new data needs to be labelled manually. Another reason is that humans can better estimate if their model requires retraining and if the improved quality still satisfies the user's needs.\\ 
Another possible trigger is a change of the infrastructure, hardware, or architectural constraints to maintain the performance and functionalities of the \ac{AI} model \cite{LopezGarcia.2020, Jackson.2018, informal_Katsiapis.2020, Wachsmuth.2012}

\subsection{Pipeline}
\label{sec:pipeline}
\begin{figure}[H]
	\centering
	\includegraphics[height=\textheight]{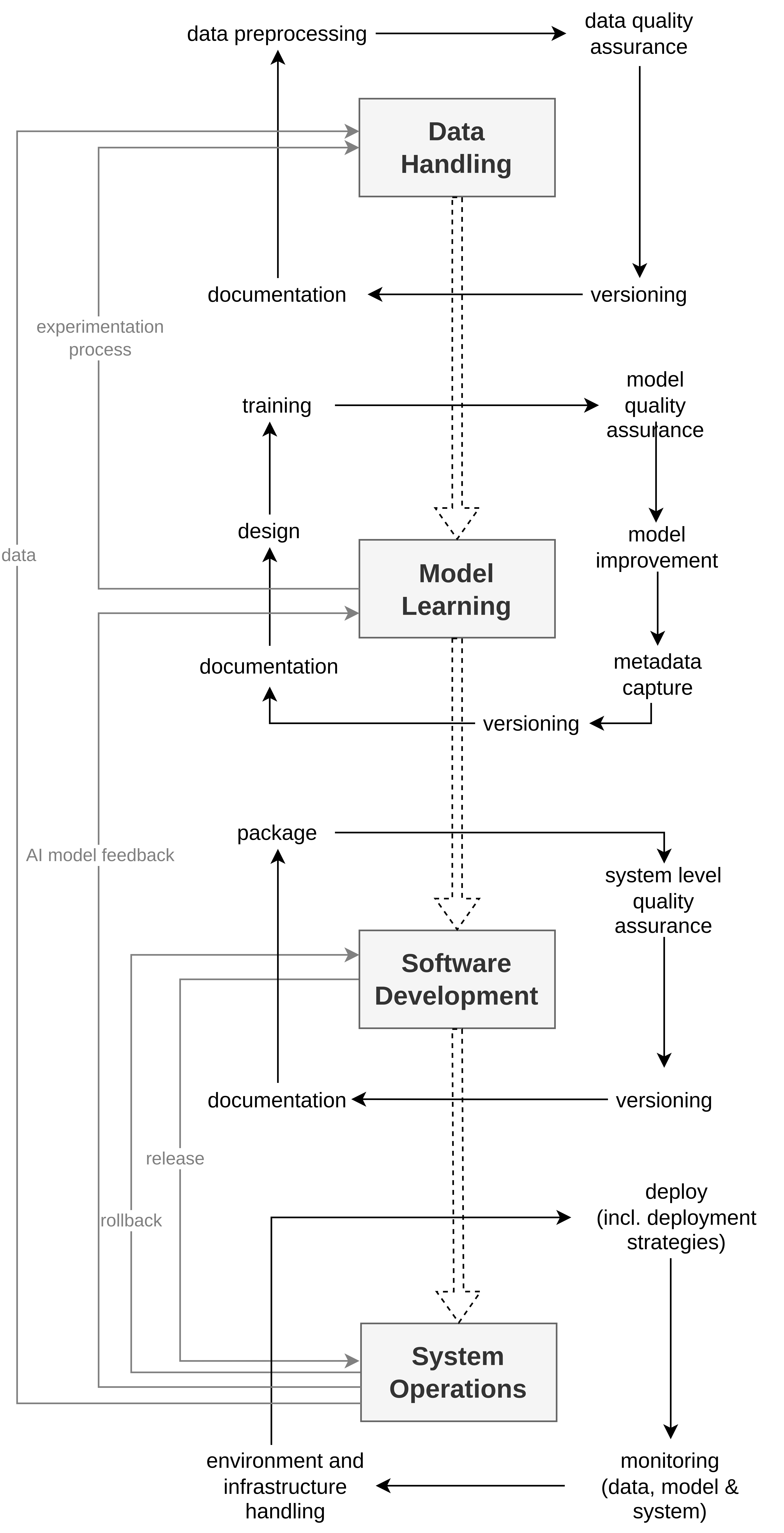}
	\caption[Continuous lifecycle pipeline for \ac{AI} applications]{Continuous lifecycle pipeline for \ac{AI} applications adapted from \cite{Lwakatare.2020} and \cite{Tamburri.2020}}
	\label{fig:pipeline}
\end{figure}

This section covers the four pipeline stages including their tasks when triggering the pipeline. The four stages are (1) \textit{Data Handling} - executing data handling, followed by the stage (2) \textit{Model Learning} - implementing the model development, the stages (3) \textit{Software Development} - building the \ac{AI} application, and(4) \textit{System Operations} - focusing on a smoothly running system and information collection in production. Figure \ref{fig:pipeline} illustrates these stages as a taxonomy that we derived from the literature and the interviews. However, Garcia et al. \cite{Garcia.2018} as well as two participants (P and T) emphasized that the pipeline and task execution strongly depends on the individual context, e.g. organizational policies for running the pipeline, and whether implementing the tasks outweigh the costs. In four out of nine participants' organizations (A, B, C and V), the pipelines are not fully automated.

As depicted in Figure \ref{fig:pipeline}, the pipeline is not linear but relies on \textbf{feedback} loops throughout the process. This is a key characteristic in agile development and allows to include continuous feedback to improve the \ac{AI} model and the collection of relevant data from production \cite{Amershi.2019}. For example, collecting data in production improves the training data set which ultimately increases model quality and efficacy \cite{informal_Khan.2018}.\\ 

Ensuring quality is an integral part of the whole engineering procedure carried out in several steps such as data, model and system-specific tests \cite{Martel.2021, Makinen.2021b}. The intention of this paper is not to give an all-encompassing picture of quality assurance techniques used for \ac{AI}. It only depicts approaches which are applicable for continuous pipelines and which we identified during the \ac{MLR}. In addition, the scope and types of tests strongly vary from implementation to implementation.

\subsubsection{Data Handling}
\label{subsec:Data}
The stage \textit{Data Handling} covers the end-to-end lifecycle of data curation. Not only allows the pipeline to handle tasks more efficiently, but also the quality of a \ac{AI} model strongly depends on the data availability, quality, and preprocessing techniques \cite{Amershi.2019, Raj.2020b, Banerjee.2020, informal_Visengeriyeva.2021}. The data pipeline manipulates the initial data via intertwined tasks, such as data preprocessing, testing, versioning, and documentation, until the data can be used for model training \cite{Raj.2020b}. A study conducted by Hummer et al. \cite{Hummer.2019} indicates that the data handling uses 7\% of the total execution time, but this time can be reduced due to parallelized computing procedures \cite{Poloskei.2020, Barrak.2021, Miao.2017, Hummer.2019, Brumbaugh.2019}. This is possible because workflows may be specified as a \ac{DAG} \cite{Poloskei.2020, Barrak.2021, Miao.2017, Hummer.2019, Stirbu.2021}.

\texttt{Data Preprocessing\\}
\label{subsub:dataPreprocessing}
Initially, data is prepared for model design and training \cite{informal_Visengeriyeva.2021}. Therefore, data collection including data injection, preparation, labelling and feature extraction needs to be executed to transform the raw data. This step is often defined in form of rules part of a script that defines how the raw data should be manipulated, transformed and compared \cite{Martel.2021, Lwakatare.2020b, informal_Visengeriyeva.2021}. Some steps may be skipped, if the data set was already preprocessed in previous iterations \cite{informal_Rosenbaum.2020}. It is essential that the data handling and transformations undertaken during data preprocessing in the pipeline are consistent with the data handling in production to avoid a training-serving skew \cite{Baylor.2017, informal_Zwebe.2021, informal_Castanyer.2021, informal_Pentreath.2019, informal_Liu.2020}. Three authors proposed to use \ac{TFX}, an end-to-end lifecycle management platform provided by Google, to avoid the training-serving skew by exporting the tasks for data transformations that are again used in production and the training and serving pipeline does not need syncing \cite{Baylor.2017, Polyzotis.2017, Olston.2017, informal_Zwebe.2021}. However, sometimes deviations in the data preprocessing pipelines is desirable because data set and its size differs, persistent data stores provide the data for training whereas data in production is non-static where data needs to be processed fast \cite{informal_Liu.2020, informal_Wilkiewicz.2019}.\\

Data can be \textbf{collected} from multiple distributed on-premise data centres, external public or private cloud storage \cite{Sangiovanni.2020, Lwakatare.2020, Banerjee.2020, Brumbaugh.2019, Raj.2020b}. The data sets may be already available (e.g., open source or internally available) or needs to be collected from multiple devices where the data may be stored in different formats, such as tabular data, logs, key-value stores or input files \cite{Amershi.2019, Junsung.2019, Brumbaugh.2019, AguilarMelgar.2021, Karlas.2020, Jackson.2018, Polyzotis.2018}. In cases where not enough data can be extracted, three interviewees (P, B and V) stated that they synthetically generate data to balance the data set. If too much data is available, the data set is reduced where participant A mentioned the risk of introducing bias.\\
The \textbf{preparation} strongly depends on the type of data \cite{Jackson.2018, Raj.2021, Sangiovanni.2020, Brumbaugh.2019}. For example, pipelines may discard incomplete or irrelevant data or outliers and noisy records \cite{Raj.2021, Amershi.2019, Junsung.2019, Rivero.2020, Nashaat.2019, informal_Saucedo.2020}, anonymize data \cite{Nashaat.2019}, (windowed/bucketed) aggregate data \cite{Brumbaugh.2019}, or decompress and resize images, of filter and tokenize text \cite{Raj.2021, Sangiovanni.2020,Breck.2019}. In addition, numeric data may be normalized via feature scaling \cite{Raj.2020, Raj.2020b}. The \ac{MLR} only extracted the z-transformation and Box-Cox transformations \cite{Banerjee.2020} for data preparation but did not extract further information on the technical details, algorithms or implementations of the data preparation tasks. \\
\textbf{Data labelling} is necessary for supervised learning as indicated by participant P and C. Therefore, each record receives a meaningful ground truth label indicating the expected output. Other learning techniques, such as reinforcement learning, use demonstrations as labelled data \cite{Amershi.2019, Junsung.2019}. To automate this process, the lifecycle
management platform ease.ml for example uses a model runner which identifies corresponding labels for input features \cite{Karlas.2020}. Unsupervised algorithms do not require labels, thus this task can be skipped \cite{Raj.2020b, Gharibi.2021}.\\
For \textbf{feature extraction} or \textbf{feature engineering}, necessary patterns need to be manually or automatically discovered and extracted, which is strongly dependent on the \ac{AI} model's context and algorithm used \cite{Raj.2021, Sangiovanni.2020, Amershi.2019, Brumbaugh.2019, Banerjee.2020, John.2020, Rivero.2020}. Automatic feature selection techniques include, e.g., particle swarm optimization \cite{Xue.2013}, recursive feature elimination, principal component analysis \cite{Jolliffe.2005}, and auto encoders \cite{Ng.2011}. For supervised learning, extracting relevant features guides the model training and identifies which features are worth exploring as input for the \ac{AI} model \cite{Karlas.2020, Banerjee.2020, John.2020, Polyzotis.2018}. One important part the pipeline should provide consistent feature extractions of the offline and online inference environment, which may be achieved by a point-in-time correctness. This is
necessary because for the training data set a vector consisting of features and associated labels is used. If features from the future are used in the vector which are not part of the labels, data leakages may occur \cite{Brumbaugh.2019, Polyzotis.2017}. \\

\texttt{Data Quality Assurance\\}
\label{subsub:DataTesting}
The pipeline runs continuously, thus, reusable components for testing data are necessary to avoid introducing bugs in the data and propagating bugs down to the model training \cite{Baylor.2017, informal_Wilkiewicz.2019, Breck.2019, Azimi.2021}. This makes it easier to identify faulty data in the beginning, before it negatively influences the model and computation resources are wasting for producing inadequate results \cite{informal_Wilkiewicz.2019, Breck.2019, Polyzotis.2018, Bachinger.2020, Stirbu.2021}. Participant R and B emphasized that the better the data quality is, the better the model results are.\\

\textbf{Data} and \textbf{feature validation}\label{dataFeatureValidation} validates batches of data. These batches can either be evaluated via a single-batch or inter-batch validation. Single-batch validation assumes that data batches collected in succession do not differ drastically and comply with a specific shape \cite{Breck.2019, Polyzotis.2018}. Thus, a stable description, such as a data schema, identifies expected features, feature types and values, and the correlation of different features \cite{Baylor.2017, Polyzotis.2017, Breck.2019, informal_Sato.2020}. Participant C also uses manually or automatically created data schemas. Information provided in this schema are the expected type, presence, valency \cite{Baylor.2017, Breck.2019, Caveness.2020}, and distribution of categorical values \cite{Polyzotis.2018, informal_Baroni.2018, informal_Wilkiewicz.2019}. Inter-batch validation identifies differences between different batches, such as training and serving data or successive training data sets \cite{Breck.2019}. Inter-batch validation identifies changes in the statistical characteristics or the encoding of feature values, such as using Boolean values instead of anticipated \textit{1} and \textit{0} \cite{Caveness.2020, Breck.2019, John.2020, Polyzotis.2018}. Statistical characteristics are discovered via calculating the distribution distance between training and serving data. Therefore, a distance threshold is calculated to identify if the deviation results in an error. The distance is calculated via two distributions including its probabilities \cite{Caveness.2020}. Other examples for calculating the statistical characteristics are Kullback-Leibler divergence, cosine similarity, or statistical goodness-of-fit tests. These approaches are implemented in \ac{TFX}\cite{Breck.2019}. Participant A and R check the right shape of data as common validation practice.\\

Data can also be validated via a \textbf{data quality} measurement which identifies how suitable the data is to meet the users' requirements \cite{Azimi.2021}. Data quality can be measured via the six dimensions of data quality, namely completeness, uniqueness, consistency, validity, accuracy and timeliness \cite{Sangiovanni.2020, Amershi.2019, Azimi.2021, Renggli.2021}. Completeness identifies how well the used fraction of the data set represents the corresponding main data set or real-world data. Uniqueness states that no duplicates occur in the data set. Consistency identifies the extend to which semantic rules of a data set are violated. Validity identifies if all data values comply with a specific format. Accuracy defines how well the data is suited and certified to execute a specific task. Timeliness describes if the data set is up-to-date for a task \cite{Renggli.2021}. Another way of identifying if the available data is sufficient to meet the user requirements is an automatic feasibility analysis proposed by ease.ml. It calculates a bayes error based on already extracted features which are used for training. The Bayes error estimator identifies the minimum error rate achievable by any classifier. Then, this error is compared to a desired target performance, such as the accuracy. If ease.ml identifies the data set as insufficient, the user can clean up the data manually based on a list of dirty examples, or collect more data \cite{AguilarMelgar.2021, Renggli.2019b, Renggli.2020, Renggli.2021}.\\

\textbf{Unit tests} can consist of tests that verify that the data ingestion works correctly \cite{Lwakatare.2020}. Unit tests can either state the input, execute a transformation and state the expected output \cite{informal_Sato.2020} or use data schemas. These tests identify if there are differences between the schema and the assumptions in the code \cite{Breck.2019}.\\

\texttt{Data Versioning\\}
\ac{AI} models require massive amounts of data for model training which needs to be stored and versioned to guarantee traceability and compliance with regulations, such as the General Data Protection Regulation \cite{Zhou.2020, Tamburri.2020}. Participant C, V and D highlighted the need for a full data lineage if important models are based on this data. Hence, not only data and its dependencies but also the data processing steps need to be versioned. In practice, however, storing the entire lineage of data is fairly impossible due to space restrictions. Thus, only the latest version is stored in participant P, T and Z's case and the previous versions are suspended.\\
The results of the \Ac{MLR} did not explicitly focus on the data storage location, such as cloud or in-house storage. Participant V and D indicated that they stored their data in-house due to the user's preference and regulations regarding for private data, such as patient records.\\

\textbf{Data} is stored via uniquely identified data snapshots or via reference to the original raw data set \cite{Derakhshan.2019, Karlas.2020, Amershi.2019, Ciucu.2019, Miao.2017b}. If all snapshots cannot be stored due to space constraints, the delta in the data set can be versioned \cite{Makinen.2021b}. Because traditional version control systems, such as Git, cannot handle the amount of data \cite{Barrak.2021, Ciucu.2019, Janardhanan.2020}, new systems and tools, such as \ac{DVC} have been introduced. \ac{DVC} can store large files due to an external storage which can be combined with Git via a lightweight metadata file including a hash to indicate the data set version and data set location. The metadata file is then tracked via Git \cite{Barrak.2021, informal_SatoMartinFowler.2019, informal_OBrien.2021}. Several space
constraints occur with edge devices where data transfer costs should be avoided. Thus, appropriate locality-aware data stores store the data lineage \cite{Rausch.2019b}.

In addition, \textbf{dependencies}, \textbf{data processing steps} and \textbf{extracted features} should be versioned. This allows to compute different versions of the data set if required. Dependencies can store the relationship between a data set which was used for training or testing and the \ac{ML} model version \cite{Lwakatare.2020b, Zhou.2020, informal_Rosenbaum.2020, Bachinger.2020}. Miao et al. \cite{Miao.2017b} and participant T suggest versioning the data processing steps including applied code and metadata \cite{Lwakatare.2020b, Amershi.2019, Stirbu.2021}. Extracted features are stored in a centralized repository, such as feature stores, where the definition, access and storage of the features is standardized \cite{informal_Google.2021, informal_Hermann.2017}. For instance, the lifecycle management platforms used by Uber (Michelangelo) \cite{informal_Hermann.2017}, Facebook (FBLearner) \cite{informal_Erb.2019} or Bighead \cite{Brumbaugh.2019} provide feature stores. This avoids to repeatedly extract feature sets where similar features may have different definitions \cite{Zhou.2020, Brumbaugh.2019, informal_Zwebe.2021}.\\

\texttt{Data Documentation\\}
\label{subsub:documentationData}
Documentation includes guidelines with concrete actions such as feature cleaning or naming conventions applicable to data files or folders \cite{Polyzotis.2018, Kronberger.2020, informal_Baroni.2018}. Documentation should be extensive enough to support auditability of \ac{AI} models. Auditability defines a reviewing process where responsibilities and potential risks associated with the usage of an \ac{AI} model are identified and root causes can be analyzed in case of a failure. Auditability for \ac{AI} still is an emerging topic that is missing generally established mechanism and regulations \cite{Bourgais.2021}. Although documentation is essential, in practice only one-third of the participants rely on manual data documentation, whereas the others do not document their data related information due to missing guidelines and software support.

\subsubsection{Model Learning}
\label{subsec:Model}
After data handling, the pipeline executes tasks associated with the \ac{AI} model learning, such as model design, model training, quality assurance and improvement. Further tasks comprise metadata capturing, versioning of the model and its dependencies, and documentation. According to Zhou et al. \cite{Zhou.2020}, this stage and its respective tasks are most essential throughout the pipeline for the continuous development of \ac{AI}.

\texttt{Design\\}
Firstly, the pipeline should support taking decisions regarding the model design, such as the appropriate hypothesis \cite{Maskey.2019, informal_Popp.2019} or regarding the selection of reusable model components which align with the problem domain \cite{Gharibi.2021, Maskey.2019, informal_Wilkiewicz.2019}. Pipelines should support context-specific decisions on the best suitable algorithm, feature selection, hyper-parameter setting, data set split, and potential reduction of the training data set \cite{Makinen.2021, Amershi.2019, Junsung.2019, informal_Wilkiewicz.2019, Spieker.2019}. The \ac{MLR} results did not reveal the concept of \ac{AutoML} within pipelines which takes over design decisions, such as finding an appropriate algorithm, model selection, data selection, and parameter tuning. Participant C illustrated that modeld.io ease the decision making process by providing \ac{AutoML}.\\

\texttt{Model Training\\}
The pipeline implementation of the model training is dependent on the architecture, distribution and amount of available computation resources \cite{LopezGarcia.2020, Poloskei.2020, Zhou.2020} and context-specific algorithms, such as unsupervised learning, deep learning, and reinforcement learning \cite{Raj.2021, Karlas.2020, Benbya.2020, John.2020, Rivero.2020, Raj.2020}. Two participants identified that \ac{TFX} and MLFlow provide the necessary implementations or support for using \ac{AI} libraries.


\texttt{\ac{AI} Model Quality Assurance\\}
\label{subsub:ModelTesting}
The pipeline for the continuous development of \ac{AI} aims to provide faultless, reliable and secure software and to guarantee a trusted decision space \cite{Raj.2021, Lwakatare.2020, LopezGarcia.2020}. Thus, fail-safe measures need to be introduced via relevant, sufficient and repeatable tests to cover all potential cases due to the continuous pipeline \cite{Fehlmann.2020, Makarov.2021, Fehlmann.2020, Baylor.2019}, which is especially important if models' decisions impact human's well-being or can cause harm \cite{Raj.2021, Makarov.2021, Arora.2019, informal_Gorcenski.2019}, as in participant D's case.\\
Due to these repetitive tests, test data sets are used more often which may lead to \textbf{overfitting}. Overfitting occurs when the model's version gets adapted throughout the pipeline cycles to pass the test \cite{AguilarMelgar.2021, Renggli.2019b, Karlas.2020, Renggli.2021, Renggli.2019}. To test the model for overfitting, three metrics may be used, such as measuring the difference between the validation and test data set \cite{Hubis.2019}, the Akaike information criteria or the Bayesian information criteria \cite{Yun.2020}.\\
However, in practice automated tests are not always possible. For instance, participant D indicated that experts had to validate the model's decisions individually. Participant P also stated that they executed quality assurance tests manually due missing configurations in the pipeline but they planed on automating the tests in the future.\\

In \ac{AI} the goal is to optimize a specific metric throughout the pipeline's lifecycles instead of simply satisfying functional requirements \cite{Zaharia.2018}. For instance, the pipeline compares statistical evaluation metrics with metrics from previous model versions \cite{Nashaat.2019, Zaharia.2018, informal_Santhanam.2019, Bachinger.2020, informal_Sato.2020}. Participant A and P endorsed to use traditional quality metrics for \ac{AI}, such as F1 scores, precision and accuracy. Another approach mentioned by participant A is a Normalized Discounted Cumulative Gain for measuring the effectiveness for their search engine model. Gerostathopoulos et al. \cite{Gerostathopoulos.2019} proposed a learnability metric specifically adapted for using it during the \ac{CI}/\ac{CD} of \ac{AI} models. Learnability is measured via five SCORE learnability facets, such as solution quality, convergence, overhead, robustness and effect. This SCORE metric was specifically adapted for using it during the \ac{CI}/\ac{CD} of \ac{AI} models \cite{Gerostathopoulos.2019, Olston.2017}. \\ 
The financial technology sector requires fair models \cite{Huang.2021, informal_Ammanath.2021}, thus Huang et al. \cite{Huang.2021} established an ethics-by-design metric for the model quality assurance in the continuous development pipeline. This metric uses approaches for feature explainability\footnote{For further information please refer to the book `Interpretable Machine Learning`, Section 5: \url{https://christophm.github.io/interpretable-ml-book/lime.html}, accessed 05.07.2021}, such as \ac{LIME}, \ac{PDP}, and \ac{SHAP}, to identify whether and how much each feature contributes to the final prediction \cite{Bourgais.2021, Yun.2020, informal_SatoMartinFowler.2019}.
In practice, interviewee R, A and B stated that they did not include checks for bias in the pipeline's model quality assurance due to their non-critical domains.\\


\texttt{Model Improvement\\}
\label{subsub:ModelImprovement}
The extracted sources from the \ac{MLR} do not identify any new pipeline-specific model improvement techniques that specifically focus on previous lifecycles. Pipeline implementations use established context and model specific improvement techniques. For instance, pipelines support model compression or pruning \cite{Hummer.2019, Karlas.2018, Boovaraghavan.2021}, model hardening \cite{Gharibi.2021, delaRuaMartinezJavier.2020, Gupta.2020b}, and hyper-parameter optimization \cite{Spell.2017, John.2020, Boovaraghavan.2021, Janardhanan.2020, informal_Duvall.2018}.\\

\texttt{Metadata Capture\\}
\label{subsub:MetadataCapture}
Metadata comprise provenance information which is necessary to govern the modelling lifecycle as well as to receive additional information on the data set and \ac{AI} model \cite{Miao.2017c, Gharibi.2021, Bachinger.2020}. For example, model specific metadata comprises relevant information on the execution and deployment of \ac{AI} models, network architecture \cite{Chard.2019}, training \cite{Lwakatare.2020, Lwakatare.2020b, Chard.2019, Vartak.2016}, instance metadata and runtime metadata \cite{informal_Liu.2020, Hummer.2019, Zhang.2020, Chard.2019}. Further metadata includes information about the model's location, registration data, and information on the start and end data \cite{Lwakatare.2020, Lwakatare.2020b, Martel.2021, Raj.2021, LopezGarcia.2020, Brumbaugh.2019}. Although feature extractors provided by several lifecycle management tools, such as \ac{TFX} automatically extract metadata \cite{Gharibi.2021, informal_Wilkiewicz.2019}.\\

\texttt{Model Versioning\\}
\label{subsub:ModelVersioning}
Model versioning not only captures version model artifacts but also model dependencies to backtrack or reproduce different model versions that quickly evolve over time \cite{Vartak.2016, Miao.2017c, Miao.2017, Miao.2017b, Li.2021, informal_Baroni.2018, Peili.2018, informal_Google.2021, Garcia.2018}. Although literature highlights the importance of model versioning, several aspects why this is not possible were mentioned by three participants (A, P and B). They  indicated that they did not store all models but only stored key model versions due to resource restraints or lack of interest in previous versions. In addition, Granlund et al. \cite{Granlund.2021} stated that they cannot store model versions due to strict guidelines regarding the patient’s data and missing isolated, in-house hardware resources. The inability to store model versions was not solved by any of the extracted pipeline tasks.\\
Model dependencies capture the relationship to related elements, such as the associated data set, source code and configuration files. This allows to recreate or load a specific model without having to rerun multiple iterations to find the appropriate model \cite{Janardhanan.2020, Lwakatare.2020, Miao.2017, Miao.2017b, Makinen.2021b}. Additionally, model versioning stores the associated log files \cite{informal_Meynard.2021} and evaluation results of a model. This allows to check whether the model versions improve continuously throughout the continuous lifecycle \cite{Lwakatare.2020, Makarov.2021, Karlas.2020, Vartak.2016}.\\

Because \ac{AI} model versioning is more complex and requires more storage capacities due to the continuous development, standard version control systems, such as Git cannot be used as model repositories \cite{informal_Breuel.2020}. Potential alternatives are container registries where images are versioned \cite{informal_Singhal.2020} or model repositories that store model versions including code, metadata, test results and dependencies \cite{Chard.2019, Stirbu.2021, informal_Moesta.2020}. During the interviews, participants proposed storage facilities of MLFlow, H2O, DataRobot and Git Large File Storage.

\texttt{Model Documentation\\}
Documentation proves that the model adheres to the regulations and restrictions and works as expected. This is required to receive certifications, which is especially important in the healthcare sector according to participant D. However, seven out of nine participants identified model documentation as a good practice however they hardly ever create and maintain an up-to-date and consistent documentation. Based on the \ac{MLR} results, software support for the model documentation during the pipeline is not established yet. \\
The \ac{MLR} however, already established the information necessary in the model documentation. For instance the documentation should outline the purpose (e.g., requirements and hypothesis) and the different methodologies to achieve the set purpose (e.g., technical decisions, such as chosen model and algorithm design) to effectively mitigate \ac{AI} related risks \cite{Stirbu.2021, informal_Haakman.2020}. Commonly, decisions made in the model creation need some rework. Thus, it is essential to document which model and test design decisions have already been tried out and which results were achieved \cite{Garcia.2018, Rivero.2020}. Additionally, documented assumptions explain the reached decisions \cite{informal_Haakman.2020}. 

\subsubsection{Software Development}
\label{subsec:Dev}
After the model is developed, it must be prepared for deployment. Therefore, the pipeline orchestrates the stage \textit{Software Development} and its related tasks, such as packaging, system-level quality assurance as well as system versioning.

\texttt{Package}
\label{subsub:package}
The build process packages the code and model logic into build artifacts which are deployed in production \cite{Vuppalapati.2020, Zhang.2020, Liu.2019, informal_Liu.2020}. The best model is selected and registered in the model registry \cite{Lwakatare.2020}.\\

Several authors propose transforming the registered model into a system-independent, deployable and hardware-optimized format \cite{Makarov.2021, Zhang.2020, Castellanos.2021}. Potential formats established in the pipeline packaging task are ONNX, SavedModel, TensorRT, \ac{PMML}, and \ac{PFA}. ONNX\footnote{For information about ONNX, please refer to \url{https://onnx.ai/}, accessed 16.07.2021} is an open, standardized format built that represents AI models by using a common set of operators and file format to allow interoperability and serialization of models \cite{Raj.2020, informal_Visengeriyeva.2021}. When the pipeline handles Tensorflow models, it is transformed to SavedModel and TensorRT format [\cite{Makarov.2021, Zhang.2020}. Another possible format is \ac{PMML}\footnote{For information about PMML, please refer to: \url{http://dmg.org/pmml/v4-4-1/GeneralStructure.html}, accessed 19.07.2021}, which describes statistical and data mining formats and transforms them into an \ac{XML} configuration file \cite{Castellanos.2021, Zaharia.2018, informal_Visengeriyeva.2021}. The \ac{PFA}\footnote{For information about \ac{PFA}, please refer to: \cite{Pivarski.2016}} is a model interchange format that provides a safe execution environment for AI algorithms \cite{informal_Visengeriyeva.2021}.\\

An essential task in this stage is to containerizes the model and its dependencies to avoid compatibility issues \cite{Chard.2019, Corbeil.2020, LopezGarcia.2020, Liu.2019, Castellanos.2021, delaRuaMartinezJavier.2020, informal_Baroni.2018, Bachinger.2020}. Extracted sources most often mention Docker \cite{Corbeil.2020, LopezGarcia.2020, Liu.2019, Castellanos.2021, Poloskei.2020, Ciucu.2019}. In addition, three interviewees (participant R, C and V) confirm using Docker extensively. Research on other approaches is sparse, however essential for increasing efficiency in cloud computing \cite{delaRuaMartinezJavier.2020}. Thus, future research suggests using \ac{VM}s, especially lightweight \ac{VM}s, type 1 hypervisors, or uni kernels also called library operating systems \cite{delaRuaMartinezJavier.2020}.

\texttt{Software-level Quality Assurance\\}
According to the paper `Hidden Technical Debt in Machine Learning Systems` \cite{Sculley.2015}, \ac{AI} models represent only a small part of the whole system landscape where many data sources and software applications interact \cite{Baylor.2017, Huang.2021, Granlund.2021, John.2020}. Software-level quality assurance should identify the correct behaviour of the whole system landscape before the system is deployed in production \cite{informal_Sato.2019, informal_Sato.2020, informal_Guo.2020}. Participants D and R agreed on the importance of this type of quality assurance whereas participant C expressed some concerns because the integration and respective testing are not model related and should be handled via a different \ac{CI}/\ac{CD} pipeline.\\

The continuity of the pipeline requires automated software-level quality assurance tests to frequently execute the tests and compare the results. For instance, integration tests check if different services work together correctly \cite{informal_Pentreath.2019}, or check whether the obtained model prediction is correctly transferred to the whole system by comparing this test to results of the model quality assurance \cite{Lwakatare.2020, informal_Meynard.2021, informal_Kent.2019}. The pipeline can automatically execute compatibility checks between interfaces and the \ac{API} endpoints \cite{informal_Windheuser, informal_SatoMartinFowler.2019}, which is essential according to participant R when AI models are integrated as microservices. Automated stress and robustness tests identify whether the whole software can perform under expected conditions \cite{Rausch.2019, Lavin.2021} by evaluating operational metrics such as throughput, latency, and resource usages \cite{informal_Sato.2020}. For example, Uber’s end-to-end lifecycle management platform Michelangelo uses an internal benchmarking system to profile certain software parts. This allows measuring how quickly inferences are run for a specific model based on real-life data \cite{informal_Guo.2020}.

However, extracted sources reveal shortcomings in the automation of available test strategies to efficiently use them during the pipeline. For instance, because user acceptance tests require human involvement \cite{informal_Pentreath.2019, Fehlmann.2020, informal.Gorcenski.2020}, Fehlmann and Eberhard \cite{Fehlmann.2020} tries to solve it via \ac{QFD}. This strategy collects customers' expectations and needs and generates a test coverage matrix where a support vector machine generates test cases \cite{Fehlmann.2020}.

\texttt{Versioning\\}
Packaged models and the respective quality assurance results of the software level are stored \cite{Granlund.2021, informal_Baroni.2018, informal_Guo.2020}. In addition, it is also necessary to version the pipeline and its associated tasks \cite{Hummer.2019, Brumbaugh.2019, Stirbu.2021}. For instance, \ac{TFX} versions the pipeline as an artifact or source code regarding the implemented pipeline tasks \cite{informal_GoogleTFX.2021, informal_Sato.2019} Additionally, ModelDB stores the pipeline as a sequence of actions in a relational database \cite{Vartak.2016}. GitHub repositories may store an Azure DevOps pipeline \cite{informal_Rosenbaum.2020}. Moreover, the used pipeline version references the associated model versions or vice versa \cite{Brumbaugh.2019}.

\texttt{Documentation\\}
Documenting information about the development stage was not handled by any paper from the literature review but was mentioned by participant T. He proposed to document the quality assurance process and associated outcome and any additional information necessary for a software release. The participant raised the research gap on software tools automatically handling and updating the documentation.

\subsubsection{System Operations}
\label{subsec:Ops}
The proposed framework's stage \textit{System Operations} handles the deployment of the \ac{AI} model into the system landscape and handles the continuous monitoring of the data, model, and system.

\texttt{Deploy\\}  
\label{subsub:deploy}
The deployment is an essential task to make the model available to others, enable collaboration and avoid knowledge silos \cite{LopezGarcia.2020, Tamburri.2020}. But before deploying the model, the first three out of four criteria need to be fulfilled.
\begin{enumerate}
\item All preceding pipeline tasks are executed successfully, such as the quality assurance test suite \cite{Zhou.2020, Lwakatare.2020, AguilarMelgar.2021, John.2020}. 
\item The pipeline identifies the best model that is not necessarily the newly trained model \cite{Raj.2021, Poloskei.2020, Olston.2017, Karlas.2018}. However, the research gap arises how to compare model versions fairly and without any bias. Not only does the pipeline need to be aware of whether the validation uses the same data but also if the evaluation remains the same \cite{Garcia.2018, Karlas.2018, Schelter.2018}. One approach to guarantee comparability is to use an unseen dataset \cite{AguilarMelgar.2021}.
\item The model fulfills user-defined deployment criteria, such as a specified increase in accuracy \cite{AguilarMelgar.2021, Rausch.2019b} or, according to participant P, a specific benchmark.
\item Some pipelines may require human involvement, such as manual validation of tests, before deploying the \ac{AI} model to production \cite{informal_Rosenbaum.2020, informal_Aronchick.2020, informal_Schruhl.2020}. This is also essential in participant P's case.
\end{enumerate}

The model is deployed on different environments, such as cloud-managed serving platforms, server disks or remote storage, that vary in their infrastructure and strongly depend on the anticipated traffic and financial resources \cite{informal_Schruhl.2020, informal?_Vadavalasa.2020, informal_Srinivasan.2021, Makinen.2021b, Spell.2017, DiazdeArcaya.2020}. Cloud-managed serving platforms have the advantage that they can scale to accommodate the need for high-performance computing, low-latency, and memory-intensive requirements \cite{Arora.2019, Castellanos.2021}. According to the interview participants, also cloud solutions are preferred.

\texttt{Deployment Strategy\\}
Continuous experimentation is a deployment strategy that allows gathering (user) feedback during runtime \cite{Corbeil.2020, Banerjee.2020, Baylor.2017}. Examples are A/B tests \cite{Corbeil.2020, informal_Singhal.2020}, canary releases \cite{Baylor.2017, Olston.2017} and shadow deployments \cite{informal_Saucedo.2020, informal_Windheuser.2020, informal_SatoMartinFowler.2019} \cite{Makinen.2021b, delaRuaMartinezJavier.2020, informal_Ettun.2019}. Participant A uses these deployment strategies to identify potential network effects. If the \ac{AI} model is used for critical decision-making in the medical sector, deployment strategies which do not influence the system behaviour, clinical performance and safety of the patients should be used. If issues are encountered, rollbacks to already packaged models are necessary as identified during the interviews (R, A and T).

\texttt{Monitoring\\}
The pipeline also monitors the \ac{AI} model in production and collects necessary information to improve the non-deterministic model over time \cite{Stirbu.2021, informal_Windheuser, Leff.2021, informal_Wilkiewicz.2019, informal.Gorcenski.2020}. Therefore, the monitoring results should be systematically mapped to the different model versions \cite{Schreiber.2014}. \\
The \ac{MLR} results reveal that the monitoring can be split up into four aspects. Firstly, the monitoring systems collect input and output data to use for future training \cite{Miao.2017c, Lavin.2021, Breck.2019, Polyzotis.2018, Bachinger.2020, Kronberger.2020, delaRuaMartinezJavier.2020, Raj.2021, Yun.2020, informal_Google.2021}.\label{modelMonitoring} In addition, this helps to quickly identify a drift in the data set due to changes in the statistical characteristics of distribution\cite{Bachinger.2020, Kronberger.2020, Yun.2020, delaRuaMartinezJavier.2020}. Secondly, monitoring tools observe the model performance to identify if the performance deteriorates when using real-life data \cite{Lwakatare.2020, Corbeil.2020, informal_Visengeriyeva.2021}. Thirdly, it collects traditional software monitoring aspects, also called \ac{KPI}s during runtime \cite{Lwakatare.2020, Corbeil.2020, Castellanos.2021}. For instance, \ac{KPI}s consist of response time, minimal latency and throughput \cite{Lwakatare.2020, Li.2021, informal_Pentreath.2019} or resource usage and network statistics, which are especially important if the \ac{AI} application is deployed as a Service \cite{Zhang.2020, informal_Baroni.2018}. Fourthly,  may be collected to identify the impact on business outcomes, such as user engagement \cite{informal_Baroni.2018, informal_Pentreath.2019, informal_SatoMartinFowler.2019}. According to participant P, operational telemetry data is essential because although the model improves, the subjective user perception of the model may diminish. To monitor the previously mentioned aspects, participants A, P and T use Google Error Logs and Postman to identify the correct behaviour and performance of \ac{API} requests.\\

\texttt{Environment and Infrastructure Handling\\}
\ac{AI} applications may be deployed in several different environments or operational stages. The environments have varying hardware, operating system and software version dependencies \cite{Hummer.2019, Brumbaugh.2019}. Although the model operates on different environments, it must provide persistent output for a specific input \cite{informal_Stumpf.2018}. If for instance, multiple computing platforms use the same \ac{AI} application, the pipeline should support cross-platform abstraction. This allows abstracting the configuration management and low-level API calls. For example, SageMaker expects a custom Docker image including predefined entry points, Watson Machine Learning requires a zip archive which includes the code for training a model. This zip archive is then deployed in a container in the cloud \cite{Hummer.2019}.

\subsection{Challenges}
\begin{figure}[H]
	\centering
	\includegraphics[height=\textheight]{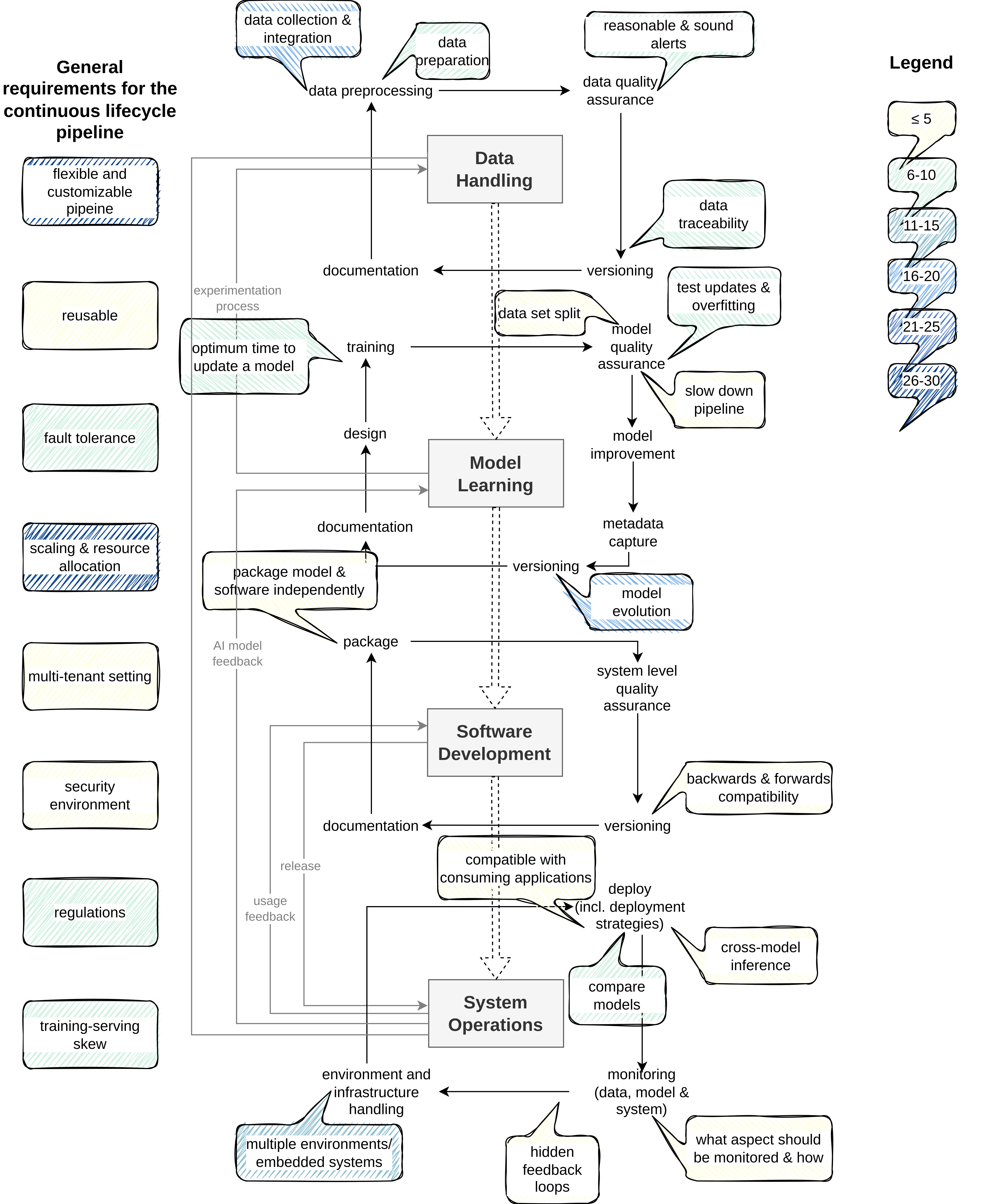}
	\caption{Challenges occurring during the implementation, adaption and usage of pipelines for the continuous development of \ac{AI} - labelled challenges based on the number of occurrences starting from yellow ($\leq5$) to dark blue ($26-30$)}
	\label{fig:pipelineChallenges}
\end{figure}

This section elaborates 25 challenges focusing on the implementation, adaption and usage of pipelines for the continuous development of \ac{AI}. Figure \ref{fig:pipelineChallenges} maps all identified challenges to general requirements or the presented framework's four stages, \textit{Data Handling}, \textit{Model Learning}, \textit{Software Development}, and \textit{System Operations}. Figure \ref{fig:pipelineChallenges} uses different colours for each challenge to indicate how often the sources state the challenge. For instance, yellow indicates that we extracted the challenge less or equal to 5 times, whereas dark blue indicates that we identified the challenge between 26 and 30 times. This section also covers a more thorough description of challenges which occurred more or equal to 16 times.

\begin{xltabular}{\linewidth}{p{1cm}p{2cm}X}
\caption{Overview of challenges regarding the implementation, adaption and usage of pipelines for the continuous development of \ac{AI}}\label{tab:challengesTable} \\

\hline

Stage & Challenge & Description  \\
\hline 
\endfirsthead

\multicolumn{3}{c}%
{\tablename\ \thetable{} -- continued from previous page} \\
\hline Stage & Challenge & Description \\ \hline 
\endhead

\endfoot

\hline
\endlastfoot
\multirow{4}{*}{\begin{sideways}\textbf{Data Handling}\end{sideways}} & Data collection \& integration & \begin{itemize}[label={--}, nolistsep, leftmargin=5pt,
            before*={\mbox{}\vspace{-\baselineskip}},after*={\mbox{}\vspace{-\baselineskip}}]
        \item comply with local data regulations \cite{Banerjee.2020, Granlund.2021}
        \item transform data from various schema regimes into universal format \cite{Amershi.2019, Renggli.2021}
        \item automatic feature labelling \cite{Polyzotis.2017, Lwakatare.2019}
        \item how to automatically identify which data source augments features \cite{Polyzotis.2017}
    \end{itemize}\\
    & Data preparation & 
 \begin{itemize}[label={--}, nolistsep, leftmargin=5pt,             before*={\mbox{}\vspace{-\baselineskip}},after*={\mbox{}\vspace{-\baselineskip}}]
    \item pipeline should identify how data preparation effects model quality \cite{Renggli.2021}
    \item cleaning raw data, the pipeline struggles to identify which effect noise or uncertainty have on the model quality \cite{Renggli.2021}
  \end{itemize} \\
 & reasonable \& sound alerts & 
 \begin{itemize}[label={--}, nolistsep, leftmargin=5pt,             before*={\mbox{}\vspace{-\baselineskip}},after*={\mbox{}\vspace{-\baselineskip}}]
    \item unclear data anomaly alerts \cite{Baylor.2017, Polyzotis.2017}
    \item correct \& reasonably strict alerts \cite{Baylor.2017, Polyzotis.2017}
    \item unclear what optimal amount of alerts is \cite{Baylor.2017}
    \item flexible \& automated data validation techniques to avoid unreasonable customization but allow for custom checks \cite{Baylor.2017}
  \end{itemize} \\
 & Data traceability & 
    \begin{itemize}[label={--}, nolistsep, leftmargin=5pt,             before*={\mbox{}\vspace{-\baselineskip}},after*={\mbox{}\vspace{-\baselineskip}}]
    \item automatic versioning methods for data sets and associated model \cite{informal_Rosenbaum.2020}
    \item store information on data governance  \cite{informal_SatoMartinFowler.2019}
  \end{itemize} \\
\hline
\multirow{6}{*}{\begin{sideways}\textbf{Model Learning}\end{sideways}}& Optimum time to update a model & \begin{itemize}[label={--}, nolistsep, leftmargin=5pt,             before*={\mbox{}\vspace{-\baselineskip}},after*={\mbox{}\vspace{-\baselineskip}}]
    \item what are signs for optimal retraining time \cite{Makinen.2021, Makinen.2021b, Banerjee.2020}
    \item pipeline ought to handle conceptual coupling between data, model and pipeline \cite{Barrak.2021}
    \item how can it handle online model retraining (pipeline is not triggered) \& model is not static artifact \cite{informal_SatoMartinFowler.2019}
  \end{itemize} \\
 & Data set split & \begin{itemize}[label={--}, nolistsep, leftmargin=5pt,             before*={\mbox{}\vspace{-\baselineskip}},after*={\mbox{}\vspace{-\baselineskip}}]
    \item how to distribute data to training, validation and test data set based on pipeline's knowledge\cite{Schelter.2018}
  \end{itemize} \\
 & test updates \& overfitting & \begin{itemize}[label={--}, nolistsep, leftmargin=5pt,             before*={\mbox{}\vspace{-\baselineskip}},after*={\mbox{}\vspace{-\baselineskip}}]
    \item test automation of self-adaptive models throughout lifecycle \cite{informal_Felderer.2021}
    \item how can the pipeline efficiently reuse tests and data set without overfitting the model \cite{Schelter.2018, Garcia.2018, Renggli.2021}
  \end{itemize}\\
 & Slow down pipeline & \begin{itemize}[label={--}, nolistsep, leftmargin=5pt,             before*={\mbox{}\vspace{-\baselineskip}},after*={\mbox{}\vspace{-\baselineskip}}]
    \item how should pipeline handle latency due to data preparation \cite{Raj.2020b}
  \end{itemize}\\
 & Model evolution & \begin{itemize}[label={--}, nolistsep, leftmargin=5pt,             before*={\mbox{}\vspace{-\baselineskip}},after*={\mbox{}\vspace{-\baselineskip}}]
    \item how can various versions of \ac{AI} models and all the dependencies be stored efficiently \cite{Lwakatare.2020, Lwakatare.2020b, Miao.2017}
    \item research identified that frequent model updates also require   corresponding changes to various software artifacts (e.g.
build files). A challenge for the pipeline is to reduce this overhead \cite{Barrak.2021}
  \end{itemize} \\ \\
\hline
\multirow{2}{*}{\begin{sideways}\textbf{Software Deployment}\end{sideways}} & Package model \& software independently & \begin{itemize}[label={--}, nolistsep, leftmargin=5pt,             before*={\mbox{}\vspace{-\baselineskip}},after*={\mbox{}\vspace{-\baselineskip}}]
    \item handle complex models that includes other models \cite{Lwakatare.2019}
  \end{itemize}\\
& Backwards \& forwards compatibility & \begin{itemize}[label={--}, nolistsep, leftmargin=5pt,             before*={\mbox{}\vspace{-\baselineskip}},after*={\mbox{}\vspace{-\baselineskip}}]
    \item how to ensure compatibility of a data schema \cite{informal_SatoMartinFowler.2019, Schelter.2018}
  \end{itemize}  \\
\hline
\multirow{6}{*}{\begin{sideways}\textbf{System Operations}\end{sideways}} & Compatible with consuming applications & \begin{itemize}[label={--}, nolistsep, leftmargin=5pt,before*={\mbox{}\vspace{-\baselineskip}},after*={\mbox{}\vspace{-\baselineskip}}]
    \item the pipeline needs to support non-standardized integration strategies for multiple consuming application (e.g. library dependency, separate service, model as data) \cite{Zaharia.2018, John.2020, Raj.2020}
  \end{itemize}\\
 & Cross-model inference & \begin{itemize}[label={--}, nolistsep, leftmargin=5pt,             before*={\mbox{}\vspace{-\baselineskip}},after*={\mbox{}\vspace{-\baselineskip}}]
    \item How can the pipeline anticipate \& prevent cross-model inferences (e.g. latency) \cite{Baylor.2017, Olston.2017}
  \end{itemize}\\
 & Compare models & \begin{itemize}[label={--}, nolistsep, leftmargin=5pt,             before*={\mbox{}\vspace{-\baselineskip}},after*={\mbox{}\vspace{-\baselineskip}}]
    \item How to identify optimum model for deployment based on stored information (when e.g. information regarding model's behaviour with non predictable production stream is not available) \cite{Makinen.2021, Miao.2017, Miao.2017b, informal_Hermann.2017, Renggli.2021}
  \end{itemize}\\
 & What aspects should be monitored \& how & \begin{itemize}[label={--}, nolistsep, leftmargin=5pt,             before*={\mbox{}\vspace{-\baselineskip}},after*={\mbox{}\vspace{-\baselineskip}}]
    \item Which information helps to improve the pipeline \cite{Gharibi.2021}
    \item How should the pipeline most efficiently integrate human interaction \cite{Granlund.2021}
  \end{itemize}\\
 & Hidden feedback loops & \begin{itemize}[label={--}, nolistsep, leftmargin=5pt,             before*={\mbox{}\vspace{-\baselineskip}},after*={\mbox{}\vspace{-\baselineskip}}]
    \item How can pipeline prohibit propagating wrong training data (e.g. bias) \cite{Amershi.2019, Breck.2019, SchleierSmith.2015}
  \end{itemize}\\
 & Multiple environments/embedded Systems & \begin{itemize}[label={--}, nolistsep, leftmargin=5pt,             before*={\mbox{}\vspace{-\baselineskip}},after*={\mbox{}\vspace{-\baselineskip}}]
    \item How can pipeline adapt to specific requirements (e.g. resources, security) of varying environments \cite{Hummer.2019, MartinezFernandez.2021, informal_Windheuser, informal_Aronchik.2020}
  \end{itemize}\\ \\
\hline
 & Flexible \& customizable pipeline & \begin{itemize}[label={--}, nolistsep, leftmargin=5pt,             before*={\mbox{}\vspace{-\baselineskip}},after*={\mbox{}\vspace{-\baselineskip}}]
    \item How can the pipeline cater to custom selected tools, services, engineering practices, libraries, frameworks and platforms in every step of the \ac{AI} lifecycle (e.g. data processing, quality control) \cite{Hummer.2019, Renggli.2019b, Spell.2017, Martel.2021, informal_Aronchick.2020, informal_Ammanath.2021, informal_Castanyer.2021} 
  \end{itemize}\\
 & Reusable & \begin{itemize}[label={--}, nolistsep, leftmargin=5pt,             before*={\mbox{}\vspace{-\baselineskip}},after*={\mbox{}\vspace{-\baselineskip}}]
    \item Pipeline ought to provide predefined templates and patterns to minimize configuration effort while maintaining flexibility \cite{Brumbaugh.2019, Hummer.2019} (Participant R)
  \end{itemize}\\
 & Fault tolerance & \begin{itemize}[label={--}, nolistsep, leftmargin=5pt,             before*={\mbox{}\vspace{-\baselineskip}},after*={\mbox{}\vspace{-\baselineskip}}]
    \item How can pipeline handle failures in its underlying execution environment due to different tools and infrastructures plugged together without interruption \cite{Hummer.2019, Baylor.2017, Arora.2019}
    \item How can the pipeline recover from intermittent failures (e.g. inconsistent data) \cite{Boag.2017}
  \end{itemize}\\
 & Scaling \& resource allocation & \begin{itemize}[label={--}, nolistsep, leftmargin=5pt,             before*={\mbox{}\vspace{-\baselineskip}},after*={\mbox{}\vspace{-\baselineskip}}]
    \item Pipeline needs to automatically adopt \& efficiently distribute resources due to the model's unpredictable complexity \cite{Martel.2021, Brumbaugh.2019, Brumbaugh.2019, delaRuaMartinezJavier.2020, Boovaraghavan.2021, informal_Sato.2019} (participant C)
    \item How can the pipeline optimize its different tasks (e.g. data processing could be optimized via declarative abstraction) \cite{Schelter.2018}
  \end{itemize}\\  
 & Multi-tenant setting & \begin{itemize}[label={--}, nolistsep, leftmargin=5pt,             before*={\mbox{}\vspace{-\baselineskip}},after*={\mbox{}\vspace{-\baselineskip}}]
    \item How can the pipeline prioritize different users all triggering the same pipeline \cite{AguilarMelgar.2021, Karlas.2018}
    \item How should the pipeline allocate resources to the tenants in a non-discriminatory manner \cite{Karlas.2018} 
  \end{itemize}\\
 & Security environment & \begin{itemize}[label={--}, nolistsep, leftmargin=5pt,             before*={\mbox{}\vspace{-\baselineskip}},after*={\mbox{}\vspace{-\baselineskip}}]
    \item How can the company-specific security environment integrate the continuous pipeline (participant Z)
  \end{itemize}\\
 & Regulations & \begin{itemize}[label={--}, nolistsep, leftmargin=5pt,             before*={\mbox{}\vspace{-\baselineskip}},after*={\mbox{}\vspace{-\baselineskip}}]
    \item How can tasks throughout the pipeline support country-specific regulations (e.g. privacy, necessary documentation, GDPR, certifications)  (participant A,T,P,R,B,V,D)
  \end{itemize}\\
 & Training-serving skew & \begin{itemize}[label={--}, nolistsep, leftmargin=5pt,             before*={\mbox{}\vspace{-\baselineskip}},after*={\mbox{}\vspace{-\baselineskip}}]
    \item Pipeline should orchestrate data processing regarding training and inference data and its specific characteristics \cite{Baylor.2017, Olston.2017, Polyzotis.2017, John.2020}
  \end{itemize}\\
 
\end{xltabular}

\subsubsection{Data Handling}
Challenges during the stage \textit{Data Handling} elaborate challenges regarding data collection and integration, preparation, data quality assurance with reasonable and sound alerts and data versioning with the respective data traceability. The most often mentioned data stage issue is the \textbf{data collection and integration} with over 16 mentioned cases. The pipeline is challenged with complying with local data regulations \cite{Banerjee.2020}. For example, the healthcare sector may require to keep the data sets within the organization. Thus, the pipeline should be able to work with restricted on-premise computation resources \cite{Granlund.2021}. In addition, participants R and D stated that the pipeline should handle data governance and ownership constraints.\\
In addition, the pipeline should automatically transform data from various schema regimes into a universal format. Additionally, the pipeline should store the data transformation and data preparation steps in a machine-readable form \cite{Amershi.2019, Renggli.2021}. Moreover, the pipeline faces the issue of automatically encoding data into features with the help of self-labelling consistent, and accurate instrumentation \cite{Polyzotis.2017, Lwakatare.2019}.\\

\subsubsection{Model Learning}
Specific challenges occurring during the stage \textit{Model Learning} elaborate challenges regarding finding the optimal time to update a model to save time and resources, model quality assurance faces the challenges to split the data set, test updated and the potential overfitting and the introduced latency that slows down the pipeline. In addition, most often mentioned was that \textbf{versioning} should depict the model evolution and ensure reproducibility \cite{informal_Saucedo.2020, informal_Srinivasan.2021}. Versioning is a static procedure, however, the pipeline needs to be capable of tracking constantly learning models and the conceptual coupling between the respective data sets, model and pipeline  \cite{informal_SatoMartinFowler.2019, Barrak.2021, Bachinger.2020, Barrak.2021, informal_Sato.2019}. Additionally, the pipeline's versioning should provide enough information to identify the reason for performance and prediction changes \cite{Yun.2020}.

\subsubsection{Software Development}
Specific challenges occurring during the stage \textit{Software Development} elaborate challenges regarding packaging model and software independently to provide scalability which was challenging for participant P. The necessary tracking provenance becomes challenging if a complex model combines several other models which were trained on different data sets. It becomes even more challenging if the training data set's transformations are highly-heterogeneous \cite{Lwakatare.2019}. The pipeline should also allow for for backwards and forwards compatibility of a data schema \cite{informal_SatoMartinFowler.2019, Schelter.2018}\\ 

\subsubsection{System Operations}
Specific challenges occurring during the stage \textit{System Operations} elaborate challenges regarding the deployment, such as ensuring that the model is compatible with the consuming application, cross-model inference and comparing models. Monitoring related challenges focus on what aspects are worth monitoring and hidden feedback loops. The challenge during the \textbf{environment and infrastructure} handling emphasizes the issue of handling multiple environments and embedded systems which was identified in 12 sources. This challenge occurs if \ac{AI} applications are deployed to the cloud, such as multiple on-premise servers, edge devices, dedicated clusters or any combination \cite{Hummer.2019, MartinezFernandez.2021, informal_Windheuser}. Because these environments handle resources and security differently, the pipeline needs to adapt to these specific requirements \cite{Hummer.2019, MartinezFernandez.2021, informal_Aronchik.2020}. For instance, edge devices provide different hardware architecture, such as arm64X or x86 or sensor setups, thus the pipeline needs to provide personalized docker containers automatically tailored to the architecture. This should allow improving hardware stability \cite{Raj.2020}.\\

\subsubsection{General Requirements for the Pipeline}
General challenges which cannot be mapped to a specific stage of the framework focus on the need of a flexible, customizable, reusable and fault tolerant pipeline. Elastic scaling is also a pipeline specific challenge because capacity changes. The required multi-tenant setting, integration into the already existing security landscape, and missing regulations pose challenges as well.\\

The pipeline has to be \textbf{flexible} and \textbf{customizable} to avoid becoming too restrictive \cite{informal_Seyffarth.2019} and to allow using preferred and established tools, services and engineering practices \cite{Hummer.2019, Renggli.2019b, Spell.2017}. Thus, pipelines need to combine quickly evolving tools, libraries, frameworks and platforms in every step of the \ac{AI} lifecycle management to meet specific requirements \cite{Martel.2021, informal_Aronchick.2020, informal_Ammanath.2021, Spell.2017, informal_Castanyer.2021}. The need for flexibility and customizability were confirmed by more than half of the participants (R, A, P, Z and C) because pipelines are very context-specific and strongly depend on the use case. Participant V uses \ac{TFX} for their \ac{AI} development and they discovered that the infrastructure handling is very difficult for them and for their customers. For example, \ac{TFX} requires Kubeflow which only works on Linux. Participant V, indicated that his customers wanted to execute the \ac{TFX} pipeline by themselves but did not have an infrastructure based on Linux.\\

The demand for computing capacities varies heavily due to the increase in the algorithms' unpredictable complexity \cite{Martel.2021, Brumbaugh.2019, Brumbaugh.2019}. Thus, elastic \textbf{scaling} is required to provide the right amount of hardware to handle capacity changes \cite{Martel.2021, Brumbaugh.2019, Brumbaugh.2019}. The pipeline needs to adapt to either horizontal or vertical scaling which is strongly dependent on the infrastructure and platform limitations \cite{delaRuaMartinezJavier.2020}. In addition, the pipeline needs to be capable of efficiently distributing these resources. This is especially important if device capabilities are heterogeneous such as differences in the \ac{CPU} complexity, memory, number of cores or bandwidth \cite{Boovaraghavan.2021, informal_Sato.2019}.

\section{Discussion}
\label{chp:discussion}
\begin{figure}[H]
	\centering
	\includegraphics[height=\textheight]{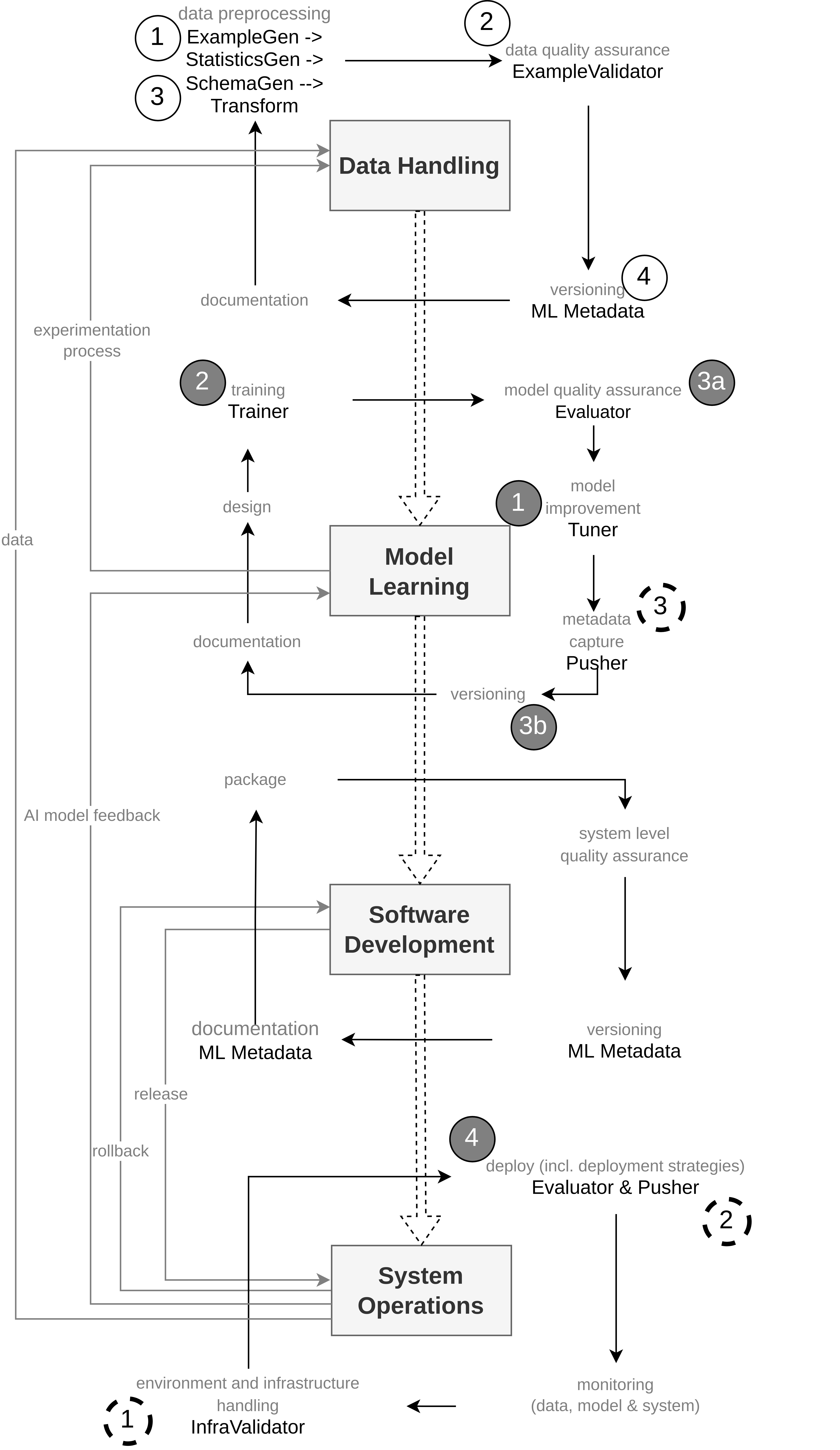}
	\caption{Map TFX's components to the proposed framework's stages and tasks; The circle with the number illustrates the execution order in TFX for each stage.}
	\label{fig:TFX}
\end{figure}

The following section provides a discussion on how to map an existing pipeline for the continuous development of \ac{AI} to the proposed framework. Whereas the proposed framework provides a thorough depiction of tasks in the lifecycle management pipeline, the implemented pipeline may vary in practice. For instance, pipelines may vary due to resource constraints, difficulties in specific implementations and because a task's benefit does not outweigh the costs. In addition, the execution order of the tasks strongly depends on the specific context, such as organizational policies for running the pipeline.\\
The most often mentioned pipeline collected via the \ac{MLR} is \ac{TFX}. \ac{TFX} combines different components to enable a flexible and customizable pipeline. An orchestrator, such as Kubeflow, manages and triggers these components. In the proposed framework, the term tasks refer to the same concept as \ac{TFX}'s components. Figure \ref{fig:TFX} illustrates the \ac{TFX}'s components mapped to the framework. Components for the data preprocessing consist of ExampleGen, StatisticsGen, SchemaGen, and Transform, which are explained in the next subsection.

\subsection{Data Handling}
\ac{TFX}'s ExampleGen component stores consumed data from different file types (e.g., csv) in an appropriate format for the following components. With the help of Apache Beam, a unified programming model for defining and executing data processing workflow, ExampleGen can read different data formats and sources. For instance, ExampleGen handles file-based data (e.g. CSV, Amazon S3, Google Cloud Storage, general text files), Messaging (e.g. Apache Kafka, Kinesis streams, JMS, Amazon Simple Queue Service), FileSystem-based data (e.g. Hadoop, Google Cloud Storage), Databases (e.g. Apache Cassandra, MongoDB, Google Cloud Datastore) and others. \ac{TFX}'s ExampleGen transforms the ingested data to customizable spans, versions and splits. A span groups together data based on context-specific characteristics, such as a day. Spans include several versions when the data is processed, resulting in a new version within the span. ExampleGen then splits each version into training and evaluation data sets.\\
StatisticsGen uses ExampleGen's output as input and calculates descriptive statistics for the data set.\\
SchemaGen uses these statistics to construct a schema automatically. The schema includes information about the (absence/presence of) data types, ranges, categories, distribution etc. Developers can modify and adapt the generated schema. \\
ExampleValidator uses the generated schema as input to validate the data set via the TensorFlow Data Validation to ensure data quality. As suggested in the \ac{MLR}, it compares the data statistics against the constructed schema, compares training and serving data to identify training-serving skews, and identifies data drifts by evaluating a series of data.\\

Opposed to the proposed framework, Transform handles feature engineering after the data validation. Transform converts the ExampleGen's data set via SchemaGen's data schema using TensorFlow Transform. Feature transformations include embedding (mapping features to a low dimensional space), vocabulary generation (convert non-numeric features into integers), value normalization (scale numeric data without distorting differences in its range), and enriching text feature (extract features from raw data, e.g. n-grams).

\subsection{Model Learning}
Opposed to the proposed framework, \ac{TFX} firstly improves the model via hyperparameter tuning, then trains the model and then validates it. 
Regarding model learning, \ac{TFX} improves the model by tuning hyperparameters via their Tuner that uses the Python KerasTuner. As input, Tuner requires the training and evaluation data set, tuning logic such as the model definition, hyperparameter search space and the Protocol buffers that include instructions for serializing structured data. \\
Trainer trains an \ac{AI} model via Python's TensorFlow API. This task uses ExampleGen's data set and the transformed features to train two models. One model gets deployed to production for inference, and \ac{TFX} uses one model for evaluation. \ac{TFX} also allows warm-starting the training by using an existing model for further training. In addition, during model training, developers can simultaneously compare multiple model runs, due to saved information in the ML Metadata Store. \\
\ac{TFX}'s Evaluator performs analysis on the training results and allows to look at the model's behaviour for individual slices of data sets. This ensures identifying well-performing models for the entire data set but poorly for a data point. This task also allows comparing models to find the optimum model that does not necessarily need to be the lastly trained model. Moreover, standard Keras metrics calculate the accuracy, precision, recall etc.

\subsection{Software Development}
\ac{TFX} does not explicitly handle any software development-specific tasks in their primary pipeline, such as packaging or system-level quality assurance. ML Metadata versions and documents software-specific information.

\subsection{System Operations}
\ac{TFX}'S InfraValidator checks whether the model complies with the production environment. To do so, it launches a sand-boxed model server based on manually configured environment details, such as type of CPU, memory, or accelerators to evaluate the compatibility between the model server binary and the model that should get deployed. \\
After successfully evaluating the model concerning the specific production serving environment, \ac{TFX}'s Pusher pushes the validated model to the appropriate deployment environment. Depending on the deployment target Pusher supports model repositories such as Tensorflow Hub, Javascript environments (tensorflow.js), native mobile applications (TensorFlow Light), or server farms (Tensorflow Serving).

\ac{TFX} versions all the artifacts produced by every component over several executions. Therefore, it uses a ML Metadata Store, a relational database that stores the properties of trained models and the respective data set, evaluation results, execution records and the provenance of data objects. Thus, the metadata store allows identifying which features impacted the evaluation metrics. It also allows only rerunning necessary components, such as skipping data processing when the developer only adapts the model hyperparameters.

\section{Threats to validity}
\label{sec:threats}
This section discusses the four possible main threats to validity according to Wohlin et al. \cite{Wohlin.2012} as well as how we mitigated these threats. In addition, the section discusses the scope of this study 
\textbf{Conclusion validity} is restricted because a qualitative approach was chosen. To improve the interviews' reliability of treatment implementation, we implemented and tested an interview plan. In addition, the semi-structured approach allowed us to ask follow-up and clarification questions to reduce misunderstandings and ensure a thorough understanding.\\
\textbf{Internal validity} may occur because only one researcher conducted the \ac{MLR}. Therefore, as proposed by Kitchenham and Charters \cite{Kitchenham.2007}, we execute a test-retest approach at the end of the initial source selection to provide internal consistency in the decisions to include or exclude the source as well as information extraction. The random sample comprised 30 papers where only two papers were excluded instead of included. This results in a consistency of 93,3\%. We calculate the Test-retests value for categorizing the data of the included sources. Therefore, we once more extracted the information from these 30 randomly selected sources. The average difference in the information extraction comprises 23\%. This value may seem high, however, some categories only had one supporting source during the test, and during the retest, we identified another source which is a 100\% increase from test to retest.\\
Regarding the \textbf{construct validity}, one may argue that the literature review is an interpretation of the meaning of the collected literature. Thus, to minimize construct validity, the selected literature resources were carefully evaluated via two quality checklists proposed by Kitchenham and Charters' \cite{Kitchenham.2007} for formal literature and Garousi et al.'s checklist \cite{Garousi.2016} for grey literature. In addition, we avoided construct irrelevance or construct under-representation as described by Messick \cite{Messick.1995} and applied to literature studies by Dellinger \cite{Dellinger.2005} via six countermeasures. (1) The selection process was rigorously documented, (2) the studies were assigned a 5-point Likert scale to identify the amount of contribution. (3) Over-representation was avoided by combining similar sources from the same authors and companies. (4) The described checklists identified unwarranted sources which were excluded. (5) Contrary findings were also included as well as (6) the search strings allowed to include many relevant study findings.\\
Regarding the \textbf{external validity}, we applied appropriate countermeasures, such as extensive search terms, including sources from academic studies as well as grey literature provided by industry, in addition to a two-stage selection process of the interview partners. Regarding the two-stage selection process, the target group consists of three different categories of interview partners to incorporate several points of view. The categories comprise people from academia and industry. Interview partners from the industry can already use a continuous end-to-end lifecycle management pipeline for \ac{AI} or are start-ups that develop \ac{AI} applications and the associated lifecycle pipeline. These interviews also helped to evaluate and extend the results obtained from the \ac{MLR} to provide a general depiction of tasks necessary for the continuous development of \ac{AI}.

The scope of this study is to cover a comprehensive framework for the continuous development of \ac{AI} models. Thus, this paper combines research on \ac{AI}, \ac{ML} and \ac{DL}. The paper does not differentiate between them because results from the \ac{MLR} treated them similarly and only the implementation of tasks in data handling or model learning slightly varies. In addition, the scope comprises different learning types, such as supervised, unsupervised and reinforcement learning. However, we did not consider special cases of learning techniques, such as federated learning or transfer learning. 
\section{Conclusion}
\label{chp:conclusion}
In this paper, we aimed to provide a comprehensive, evidence-based foundation of established research on pipelines for the continuous development of \ac{AI}. Thus, the main goal was to systematically identify relevant conceptual ideas, as well as to synthesize and structure the research in the area. To achieve this goal, we extracted 151 relevant formal and informal sources via a \acf{MLR} and we executed nine semi-structured interviews \cite{steidl_monika_2022_5902776}. Based on this information, we identified and compared five primarily used terms, such as \ac{DevOps} for \ac{AI}, \ac{CI}/\ac{CD} for \ac{AI}, \ac{MLOps} describing an extension of \ac{DevOps}, the term end-to-end lifecycle management to describe the continuous execution of development and deployment tasks, and \ac{CD4ML} describing the technical implementation of \ac{MLOps}.\\
The paper also investigated potential main \textbf{triggers}, such as feedback and alert systems, orchestration service with a scheduled time, traditional repository updates, and manual triggers.\\
These triggers start the execution of the \textbf{pipeline}, which consists of four stages: (1) \textit{Data Handling}, (2) \textit{Model Learning}, (3) \textit{Software Development} and (4) \textit{System Operations}. The stage \textit{Data Handling} comprises the repetitive end-to-end lifecycle of data-related tasks, such as pre-processing, quality assurance, versioning, and documentation. The stage \textit{Model Learning} uses the output of the data handling and illustrates the tasks associated with the model development, such as model design, training, quality assurance, model improvement metadata capture, versioning, and documentation. After the the pipeline handles the model learning, the stage \textit{Software Development} prepares the model for deployment via packaging, software level quality assurance, and system versioning. The final stage \textit{System Operations} deploys the \ac{AI} model to a specific environment via different deployment strategies and monitors the system.\\

Furthermore, the paper maps 25 potential \textbf{challenges} regarding the implementation, adaption, and usage of pipelines for the continuous development of \ac{AI} to the previously mentioned four stages. \textit{Data Handling} related challenges encompass data collection and integration. \textit{Model Learning} related challenges identify that versioning should depict the model evolution and ensure reproducibility. \textit{Software Development} requires the pipeline to provide backward and forward compatibility of the model and data schema. The stage \textit{System Operations} needs to handle multiple environments where cross-model inferences may occur. Ultimately challenges resulting from the pipeline requirements, such as the need for a flexible, customizable, and scalable pipeline are elaborated.\\

\label{sec:FutureContribution}
Regarding \textbf{future work}, we plan to further explore, evaluate and compare the pipeline concepts by prototypical implementations of demonstrators using available platforms that provide a pipeline for the continuous development of \ac{AI}. Additional candidate platforms are, for instance, MLFlow, Uber's lifecycle platform for \ac{AI}, called Michelangelo, ModelDB, or Facebook's FBLearner. The identified tasks and resulting taxonomy in this paper may then be compared with the already available tasks supported by the mentioned platforms. In addition, challenges for implementing, adapting, and using the pipeline for the continuous development of \ac{AI} can also be tracked during the implementation and demonstration to complement our findings from the literature.\\

\textbf{Acknowledgement:} This work was supported by the Austrian Research Promotion Agency (FFG) in the frame of the project ConTest [888127] and the COMET competence center SCCH/INTEGRATE [865891, 892418]. We also thank all interview participants for their valuable input and feedback.




\newrefcontext
\printbibliography[notkeyword=formal, notkeyword=informal, title=List of General References]
\newrefcontext[labelprefix=F]
\printbibliography[keyword=formal, title=List of Formal References]

\newrefcontext[labelprefix=I]
\printbibliography[keyword=informal, title=List of Informal References]

\end{document}